\documentclass[]{aiaa-tc}

\usepackage{varioref}
\usepackage{threeparttable}
\usepackage{dcolumn}
  \newcolumntype{d}{D{.}{.}{-1}}
\usepackage{nomencl}
  \makeglossary

\usepackage{caption}
\usepackage{subcaption}
\usepackage{fancyvrb}
  \fvset{fontsize=\footnotesize,xleftmargin=2em} 
\usepackage{lettrine}
\usepackage[colorlinks]{hyperref}
\usepackage{graphicx}
\usepackage{dcolumn}
\usepackage{bm}
\usepackage{color, rotating, overpic}
\usepackage{amssymb,amsmath,amsbsy}
\usepackage{dsfont}

\definecolor{blue}{rgb}{0, 0.4470, 0.7410}
\definecolor{red}{rgb}{0.8500, 0.1250, 0.0480} 
\definecolor{green}{rgb}{0.4660, 0.6740, 0.1880}

\newcommand{\eq}{Eq.~}

\newcommand{\fig}{Fig.~}

\newcommand{\etal}{{\it{et al}.}}
\def\ip<#1,#2>{\left\langle #1,#2\right\rangle}

\usepackage{draftwatermark}
\SetWatermarkText{\sf }
\SetWatermarkScale{1.8}

\usepackage{cite}

\usepackage{amsthm}
\newtheoremstyle{dotless}{}{}{\normalfont}{}{\bfseries}{}{ }{}
\theoremstyle{dotless}

\usepackage{multirow}
\usepackage{rotating}

\graphicspath{{./figs/}{./figs_cylinder/}{./figs_channel/}}

\title{Modal Analysis of Fluid Flows: \\ Applications and Outlook} 

\author{
  Kunihiko Taira\footnote{Associate Professor, Mechanical and Aerospace Engineering (UCLA), Mechanical Engineering (FSU), Associate Fellow AIAA.}\\
  {\normalsize\itshape University of California, Los Angeles, CA 90095, USA}\\
  {\normalsize\itshape Florida State University, Tallahassee, FL 32310, USA}\\
  \and
  Maziar S. Hemati\footnote{Assistant Professor, Aerospace Engineering and Mechanics, Senior Member AIAA.}\\
  {\normalsize\itshape University of Minnesota, Minneapolis, MN 55455, USA}\\
  \and
  Steven L. Brunton\footnote{Associate Professor, Mechanical Engineering, Member AIAA.}\\
  {\normalsize\itshape University of Washington, Seattle, WA 98195, USA}\\
  \and  
  Yiyang Sun\footnote{Postdoctoral Associate, Aerospace Engineering and Mechanics, Member AIAA.}\\
  {\normalsize\itshape University of Minnesota, Minneapolis, MN 55455, USA}\\
  \and  
  Karthik Duraisamy\footnote{Associate Professor, Aerospace Engineering, Member AIAA.}\\ 
  {\normalsize\itshape University of Michigan, Ann Arbor, MI 48109, USA}\\
  \and  
  Shervin Bagheri\footnote{Associate Professor, Mechanics.}\\
  {\normalsize\itshape Royal Institute of Technology, Stockholm, 10044, Sweden}\\
  \and
  Scott T. M. Dawson\footnote{Assistant Professor, Mechanical, Materials, and Aerospace Engineering (IIT), Postdoctoral Research Scholar, Graduate Aerospace Laboratories (Caltech), Member AIAA.}\\
  {\normalsize\itshape Illinois Institute of Technology, Chicago, IL 60616, USA}\\
  {\normalsize\itshape California Institute of Technology, Pasadena, CA 91125, USA}\\
  \and
  Chi-An Yeh\footnote{Postdoctoral Research Scholar, Mechanical and Aerospace Engineering (UCLA), Mechanical Engineering (FSU), Member AIAA.}\\
  {\normalsize\itshape University of California, Los Angeles, CA 90095, USA}\\
  {\normalsize\itshape Florida State University, Tallahassee, FL 32310, USA}\\
 }
 
 \AIAApapernumber{YEAR-NUMBER}
 \AIAAconference{Conference Name, Date, and Location}
 \AIAAcopyright{\AIAAcopyrightD{YEAR}}

\begin{document}

\maketitle


\section{Introduction}

The field of fluid mechanics involves a range of rich and vibrant  problems with complex dynamics stemming from instabilities, nonlinearities, and turbulence.  The analysis of these flows benefits from having access to high-resolution spatio-temporal data that capture the intricate physics.  With the rapid advancement in computational hardware and experimental measurement techniques over the past few decades, studies of ever more complex fluid flows have become possible.  While these analyses provide great details of complex unsteady fluid flows, we are now faced with the challenge of analyzing vast and growing data and high-dimensional nonlinear dynamics representing increasingly complex flows.  

\begin{figure}
 \begin{center}
 \includegraphics[width=0.4\textwidth]{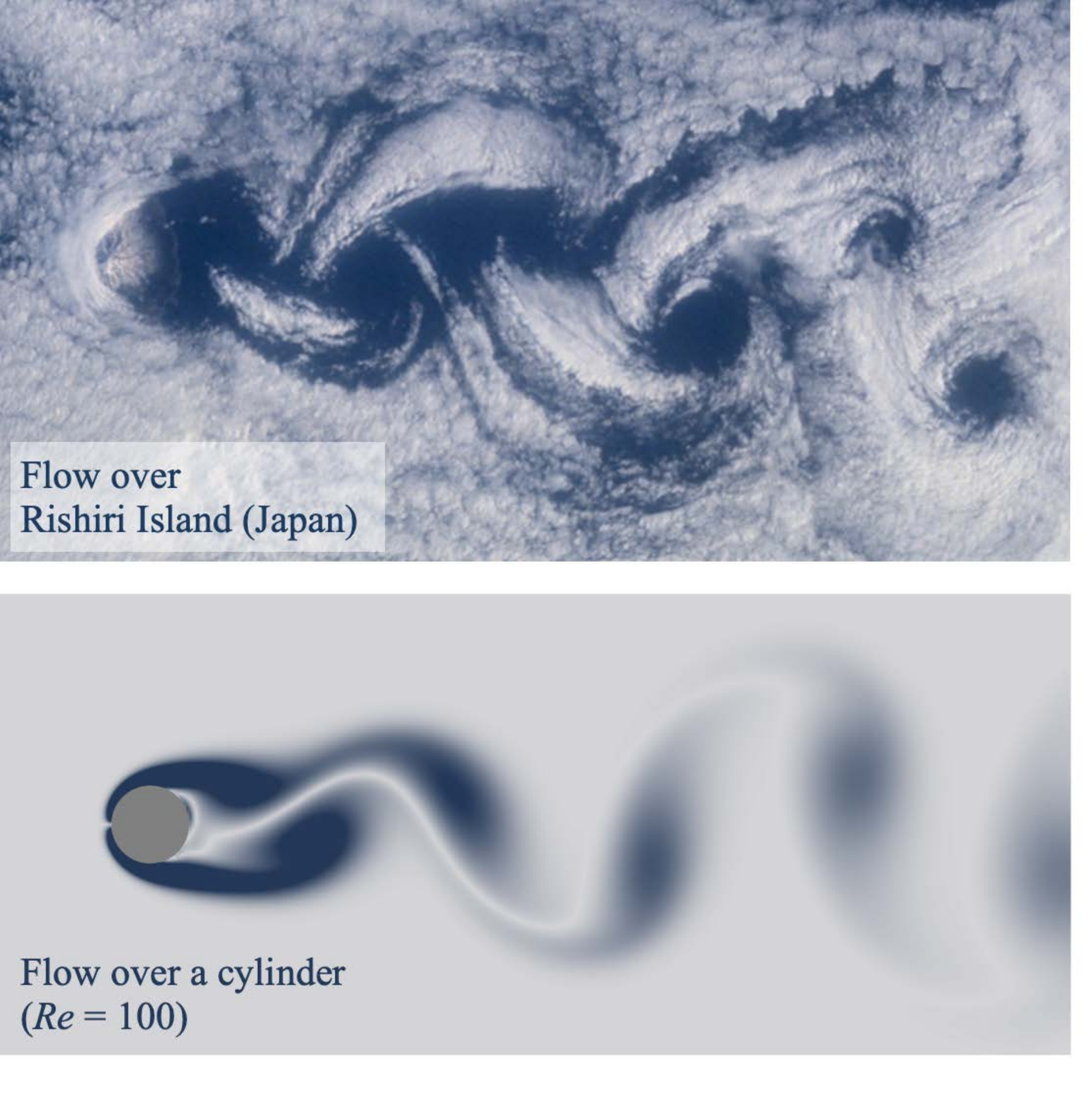}
 \caption{The von K\'arm\'an vortex street generated by the Rishiri island of Hokkaido, Japan (top, photo from NASA, 2001; STS-100).  This wake produced at high Reynolds number shares great similarity with the cylinder wake at low Reynolds number (bottom).}
 \label{fig:island}
 \end{center}
\end{figure}

Although the analysis of these complex flows may appear daunting, the fact that common flow features emerge across a wide spectrum of fluid flows or over a large range of non-dimensional flow parameters suggests that there are key underlying phenomena that serve as the foundation of many flows.  The emergence of these prominent features, including the von K\'arm\'an vortex shedding and the Kelvin--Helmholtz instability, provides hope that a lot of the flows we encounter share low-dimensional features embedded in high-dimensional dynamics.  Shown as an example in \fig \ref{fig:island} is a photograph taken from the Space Shuttle (STS-100) of the von K\'arm\'an vortex street generated by the Rishiri Island of Japan, whose wake is visualized by the clouds. Let us compare this image with the two-dimensional low Reynolds number flow over a circular cylinder shown in the same figure.  The striking similarity between these two flows suggests the existence of spatial features that capture the essence of the flow physics.  In this work, we present modal analysis techniques to mathematically extract the underlying flow features from flow field data or the flow evolution operators.

In addition to flow \emph{analysis}, modal decomposition techniques can also be used to facilitate reduced-order flow \emph{modeling} and \emph{control}.  Indeed, modal decomposition techniques offer a powerful means of identifying an effective low-dimensional coordinate system for capturing dominant flow mechanisms.  The reduction of the system order corresponds to the choice of an appropriate (reduced basis) coordinate system to represent the fluid flow.  This concept has implications for nearly every ensuing modeling and control decision.  A linear subspace to describe the flow, for example, obtained via proper orthogonal decomposition, is the most common choice for a low-dimensional basis.  After the choice of coordinate system, there are two main distinctions in modeling procedures: depending on (1) whether or not the model is physics-based or data-driven, and (2) whether or not the model is linear or nonlinear.  Further discussion of approaches for modeling and controlling fluid flows will be explored throughout the paper, with particular emphasis in the Outlook section.

The present paper is one of the products from the AIAA Discussion Group on Modal Decomposition Methods for Aerodynamic Flows (2015-2018) organized under the support of the AIAA Fluid Dynamics Technical Committee.  The goal of this discussion group was to provide an educational service to the non-specialist who seeks to gain greater insights from fluid flows with modal decomposition and analysis methods.  Since the discussion group started in 2015, invited sessions were organized at the 2016, 2017, and 2018 Aviation Meetings.  The insights gained from various discussions group activities led to the previous overview paper on modal analysis methods that focused on a broad review at a fundamental level [\citen{Taira_etal:AIAAJ17}].  The modal analysis methods described at length in the previous overview paper make up an arsenal of versatile tools relevant for extracting concise and interpretable characterizations of complex spatio-temporal flow field data, or the operators that generated them.  In the first overview paper, we covered the proper orthogonal decomposition (POD) [\citen{Holmes96,Berkooz1993}], balanced proper orthogonal decomposition (BPOD) [\citen{Rowley-ijbc05,Ilak:PF08}], dynamic mode decomposition (DMD) [\citen{Schmid:JFM10, Rowley:JFM09, Tu:JCD14, Kutz2016book}], Koopman analysis [\citen{Rowley:JFM09, Mezic:ARFM13}], global stability analysis [\citen{Theofilis:PAS03, Theofilis:ARFM11}], and resolvent analysis [\citen{Trefethen:Science93}, \citen{McKeonPoF13}].

While the first overview paper~[\citen{Taira_etal:AIAAJ17}] focused 
on the \emph{how to} aspects of modal analysis methods,
the present effort aims to demonstrate how the outputs of these modal analysis techniques can be {\it interpreted} to elucidate physical insights.
Indeed, a blind application of a modal analysis method is rarely a worthwhile endeavor.
The compilation of the present paper is primarily motivated by the fact that the true power of modal analysis techniques
in practice stems from the user's ability to appropriately apply these methods and to interpret the associated outputs.
In addition to this second overview paper, we have assembled a number of modal analysis papers to form a Special Section in the AIAA Journal to document the efforts of the discussion group.  The purpose of this special section of the present issue is to serve as an educational support to provide sufficient guidance on how modal analysis techniques can be used to extract useful and relevant information from otherwise complex flow physics.
We assume here that readers of this paper are familiar with the basics of modal decomposition and analysis techniques covered in the first overview paper [\citen{Taira_etal:AIAAJ17}].  For this reason, we do not reintroduce the algorithms for performing modal analysis, but rather focus on presenting the interpretations of the results from these analyses.

In what follows, we select a few applications of modal analysis techniques on a number of canonical flows that capture fundamental flow features in many engineering and scientific settings.
In particular, we consider examples of cylinder wakes, wall-bounded flows, airfoil wakes, and cavity flows.  A short summary of the topics covered in this paper is compiled in Table~\ref{table:overview}.
The examples within this overview paper are mostly of a computational nature.  However, this should not discourage users from employing relevant modal analysis techniques to analyze experimental fluid flows. 
Experimental datasets introduce a unique set of challenges (e.g.,~noisy and band-limited data) that must be considered carefully when using data-driven analysis methods. In this overview, we highlight some of the prevailing challenges for analyzing experimental data and point to best practices and relevant references where applicable.
We also note that the references highlighted in the examples below are not necessarily the first to perform a modal analysis of the associated flows.  
Recent citations are provided in the paper to serve as an educational guidance and chosen in hopes of facilitating interested readers to dive further into past literature.
Emphasis of the discussions is placed on how different modal analysis techniques can {\it complement} one another to reveal different characteristics of the flow, build reduced-order models, and provide guidance for flow control.  Towards the latter part of the paper, we offer an outlook on modal analysis within fluid mechanics.

\begin{table}[t]
{\small
    \centering
    \begin{tabular}{llp{4.5in}} \hline
        & Sections              & Keywords \\ \hline \hline
    I.  & Introduction          &  \\
    II. & Cylinder wakes        & POD, DMD, global stability analysis, 
                                    flow modeling, Galerkin projection, SINDy \\   
    III.& Wall-bounded flows    &  POD, BPOD, DMD, global stability analysis, resolvent analysis, flow modeling, Galerkin projection, flow control \\
    IV. & Airfoil wakes         & POD, DMD, global stability analysis, resolvent analysis,
                                    parabolized stability analysis, flow control \\
    V.  & Cavity flows          & POD, DMD, global stability analysis, resolvent analysis,
                                    flow control, aircraft application \\
    VI. & Outlook               & Superposition, sparse \& randomized algorithms, machine learning, reduced-order models, closure, hyper-reduction \\ \hline
    \end{tabular}
    \caption{An outline of the present paper.}
    \label{table:overview}
}
\end{table}

\section{Cylinder wakes}
\label{sec:cylinder}

Flow over a circular cylinder is one of the most fundamental flows in fluid mechanics for its relevance in engineering settings and for capturing the essential features of bluff body flows  [\citen{Zdravkovich97, Zdravkovich03}].  
For these reasons, there have been a tremendous amount of analyses performed on various aspects of cylinder flows, including its wake [\citen{Karman:GN11, Roshko:NACA1191, Coutanceau:JFM77a, Coutanceau:JFM77b, Williamson:JFS88, Williamson:ARFM96}], aerodynamic forces [\citen{Roshko:JFM61}], stability [\citen{Noack:JFM94, Giannetti:JFM07, Behara:PF10}], compressibility [\citen{Canuto:JFM15}], fluid-structure interactions [\citen{Williamson:ARFM04, Sarpkaya:JFS04}], and flow control [\citen{Tokumaru:JFM91, Kim:PF05, Munday:PF13, Taira:JFM18}].
Over the past two decades, modal analysis techniques have played crucial roles in uncovering additional insights into the cylinder wake dynamics. 
While we cannot provide a complete review of modal analysis performed on cylinder flows, we discuss some modal analysis studies that describe the wake dynamics and suppress unsteadiness with flow control using mode-based reduced-order models.

Over a wide range of Reynolds numbers, cylinder flows exhibit the distinct K\'arm\'an shedding wake, even under the presence of spanwise instabilities and turbulence [\citen{Zdravkovich97, Zdravkovich03}], as shown in \fig \ref{fig:island}.  The fact that K\'arm\'an shedding is identifiable from simple visual inspection suggests that such flow structures can represent the flow field in a low-dimensional manner.  In other words, these flows can be compressed to these dominant flow features. It is worth mentioning that the cylinder wake provides an ideal setting for developing and testing 
modal analysis techniques.  Note that additional complexity in the flow is often desired to test modal analysis techniques for many applications.  Nonetheless, cylinder flow serves as an attractive initial test bed for development and for educational purposes.  Beyond admitting low-dimensional dynamic representations,
cylinder wakes have been widely studied and numerous investigations 
have been documented in the scientific and engineering literature.  Furthermore, data from both physical experiments and numerical simulations
are relatively simple to acquire and reproduce.  The results of various modal analysis techniques applied to cylinder wake are often among the easier results to interpret, and thus provide a convenient entry point for developing intuition around these methods.  For all these reasons, we begin by introducing various modal analysis approaches within the context of the cylinder wake.  The first portion of this discussion will be devoted to guiding the readers through the process of building intuition for each modal analysis technique and for developing an appreciation for how to interpret the outputs of such analyses.  The ensuing sections will focus on more advanced applications, highlighting recent efforts on uncovering flow physics associated with the cylinder wake.

\subsection{Proper orthogonal decomposition}

Let us first consider the data-based modal analysis of cylinder flow. Data-based techniques only need the flow field data obtained from numerical simulation or experimental measurements and do not require knowledge of the governing dynamics.  
In particular, we consider the proper orthogonal decomposition (POD), which can extract modal contents from a collection of snapshot data.  
The term {\it snapshot} is used in modal analysis to refer to flow field data collected at an instance in time.  Before performing POD of the snapshot data, each flow field data at an instance in time needs to be formatted into a column vector.
Details on formatting the snapshot to perform the POD analysis (as well as the DMD analysis) can be found in the Appendix of the first overview paper [\citen{Taira_etal:AIAAJ17}].  
Although we make exclusive use of the snapshot POD method in the present manuscript, analytical POD methods are also commonly employed for flow analysis and modeling. 

If the velocity field is analyzed with POD, the modes $\boldsymbol{\varphi}_i(\boldsymbol{x})$ optimally capture the kinetic energy of the unsteady flow field and the eigenvalues $\lambda_i$ represent the amount of kinetic energy held by each mode.  That means that the POD analysis finds the best set of spatial modes to extract as much kinetic energy as possible in the flow field over time. These POD modes are orthogonal to each other ensuring the optimality of extracting kinetic energy by each individual modes.  In a mathematically abstract sense, we can consider the POD analysis to be fitting a low-dimensional ellipsoid to the given data.

\begin{figure}
\centering
   \includegraphics[width=0.925\textwidth]{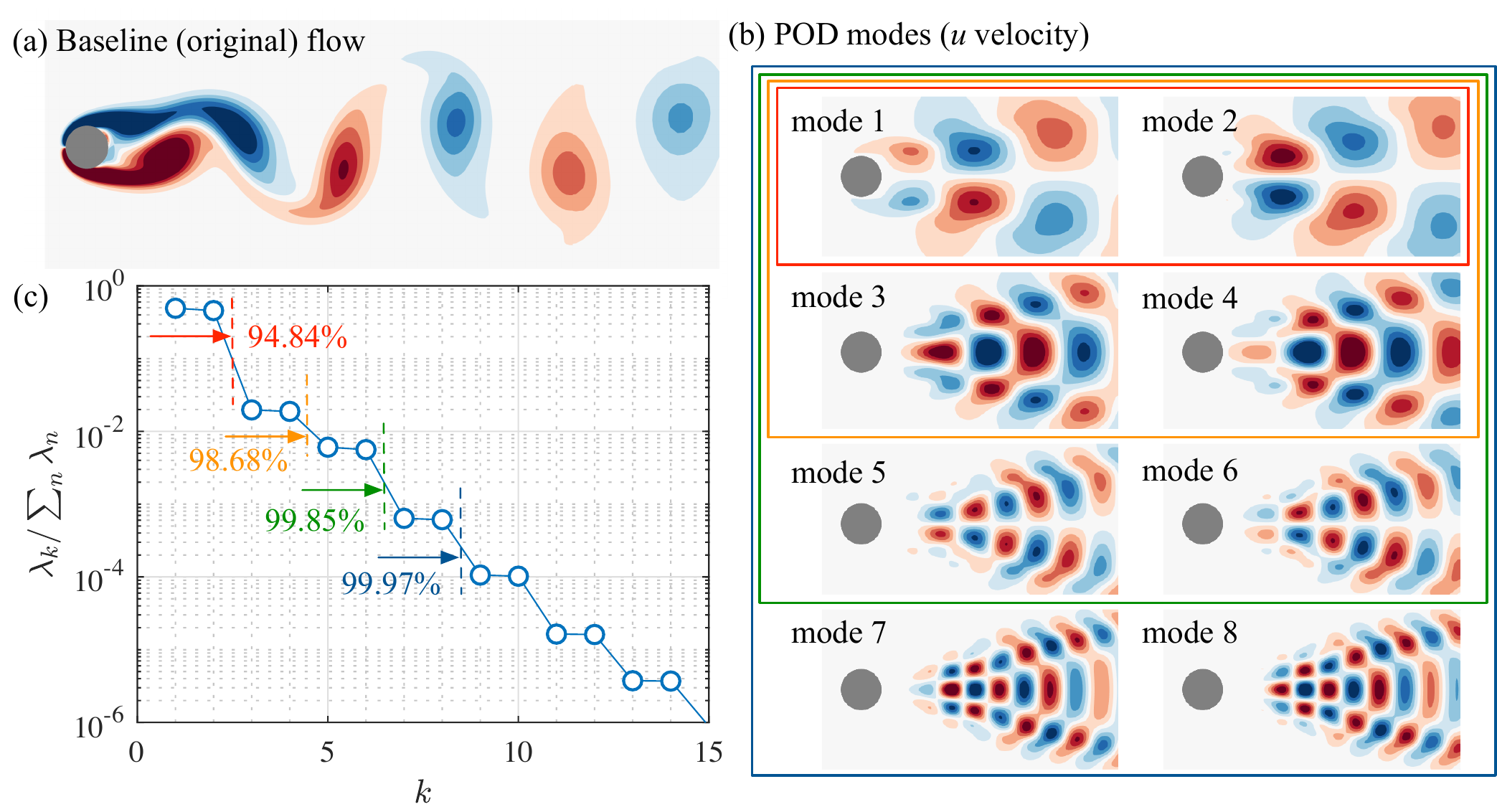}
   \caption{POD analysis of cylinder flow.  (a) Original flow field under study (vorticity shown).  (b) First 8 dominant POD modes.  (c) Amount of kinetic energy of unsteadiness captured by the POD modes.}
   \label{fig:cylinder_POD}
\end{figure}

Let us demonstrate the use of POD on the two-dimensional unsteady laminar flow over a circular cylinder at a diameter-based Reynolds number of $Re = 100$. The cylinder flow analyzed here is obtained from direct numerical simulation using the immersed boundary projection method [\citen{Taira:JCP07, Colonius:CMAME08}].  An instantaneous snapshot of the cylinder flow is shown in \fig \ref{fig:cylinder_POD}, exhibiting the von K\'arm\'an vortex shedding.  For the POD analysis, we collect 325 snapshots of the flow field over 8 shedding periods.  The data is compiled into a data matrix, upon which the snapshot POD [\citen{Sirovich:QAM87}] is applied.  The snapshot based method enables us to perform the decomposition in a computationally tractable manner when the dimension of an individual snapshot is much larger than the total number of snapshots [\citen{Sirovich:QAM87, Holmes96}].  In performing the POD here, we first subtract the mean from all snapshots, so that we can focus on modal structures associated with fluctuations.  The extracted spatial POD modes $\boldsymbol{\varphi}_i(\boldsymbol{x})$ capture regions where fluctuations appear in the flow.  Since this cylinder flow example is a periodic flow, these spatial modes appear in pairs.  This also suggests that the modes are based on advective physics with oscillator-type dynamics.

The POD analysis reveals that the fluctuations in the flow field can be captured well with only a small number of mode pairs as illustrated in \fig \ref{fig:cylinder_POD}.  The first 2, 4, and 6 modes capture $94.84\%$, $98.68\%$, and $99.85\%$, respectively, of the flow fluctuations in terms of the kinetic energy.  With 8 modes, this percentage reaches $99.97\%$, which is essentially $100\%$.  This means that the high-dimensional flow field can be accurately expressed with only 6 or 8 spatial modes, suggesting the possibility for significant compression of the flow field data.  
That is, we reduce the representation of the flow field from the number of grid points (times the number of flow variables) to merely the number of POD modes.  The mode shapes associated with the dominant POD modes reveal the dominant energetic spatial structures in the flow.  Interestingly, both POD modes 1 and 2 possess a top-down asymmetry, indicating that the dominant energetic structures are associated with the asymmetry of the K\'arm\'an wake.  As it will be discussed in a latter section, these POD modes can serve as a basis to construct a reduced-order model that describes the dynamics of the flow.  One of the important properties of the POD modes is the orthogonality of the modes (i.e., $\left< \boldsymbol{\varphi}_i, \boldsymbol{\varphi}_j \right> = \delta_{ij}$), which is attractive for developing sparse reduced-order representation of the flow dynamics.  

The above 8 POD modes can capture the flow field very well for the given data.  However, if the flow is perturbed and deviates away from the original flow, additional modes may be needed to represent the perturbed flow.  To better capture the perturbed flow, POD analysis may be repeated with the perturbed flow field data or alternative techniques such as the Balanced POD analysis [\citen{Rowley-ijbc05}] may be utilized (although an adjoint simulation is needed for the latter case).  We should keep in mind that the modes extracted from the input flow field data are optimally determined for the provided data and may not be so for the perturbed flows.  The modes may deform when the flow is under the influence of perturbation or actuation.  This is an important point to remember if modal analysis is to be extended or mode-based models are applied to perform flow control.

\subsection{Dynamic mode decomposition}

\begin{figure}
    \centering
    \includegraphics[width=0.99\textwidth]{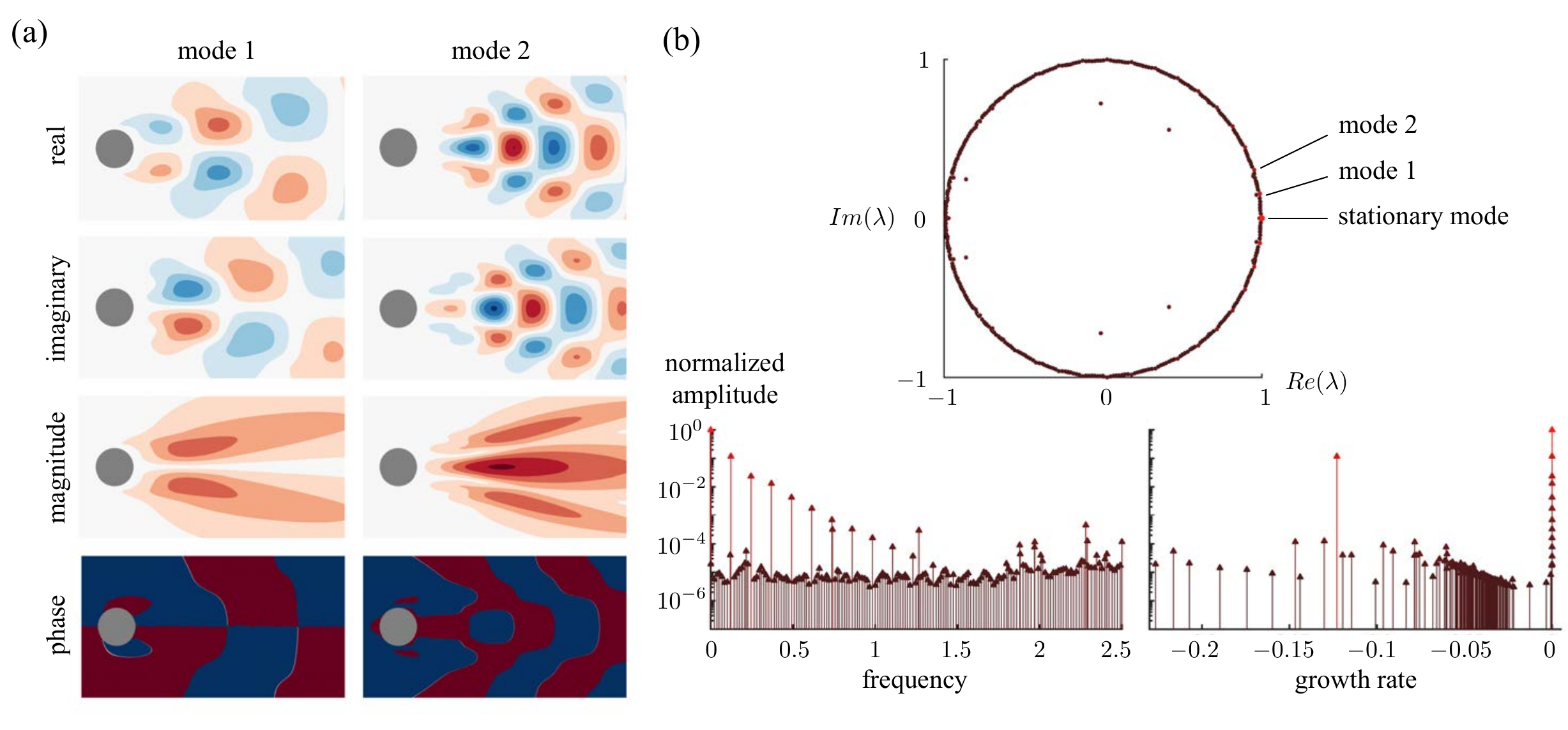}
    \caption{DMD analysis of cylinder flow.  (a) The first and second DMD modes with their real, imaginary, magnitude, and phase distributions.  (b) DMD eigenvalues representing the growth rates and frequencies.}
    \label{fig:dmd_cylinder_compiled}
\end{figure}

We now consider the second data-based approach, the DMD analysis, to study the periodic cylinder wake.  For DMD analysis, the mean subtraction from the snapshots is not necessary, unlike the POD analysis.
The findings from the DMD analysis of cylinder wake is shown in \fig \ref{fig:dmd_cylinder_compiled}.  The dominant mode that arises from the DMD analysis corresponds to a static mode (i.e.,~DMD eigenvalue $\lambda=1$), which is the mean flow.  The first two rows of \fig \ref{fig:dmd_cylinder_compiled}(a), report the real and imaginary parts of the first two oscillatory DMD modes, while in the last two rows we report the magnitude and phase of each of these modes.  Note that oscillatory modes appear in complex-conjugate pairs.  For brevity, only one element of this pair is plotted.  Consider the first oscillatory mode, visualized in different ways within the first column of \fig \ref{fig:dmd_cylinder_compiled}(a).  From both the real/imaginary and the magnitude/phase representations, it is evident that this mode captures the top down asymmetry associated with the K\'arm\'an vortex shedding, which is consistent with the POD analysis.  This is not a coincidence, which will be described shortly.  The second oscillatory mode is displayed in the second column.  The magnitude/phase plots facilitate the distillation of these physical insights.  The magnitude plots clearly reveal the active regions of each mode, whereas the phase plot displays the relative phase between spatial regions.

Let us compare the results from DMD and POD.  Unlike the oscillatory DMD modes, all POD modes are real-valued.  Since the data here is taken from a limit cycle oscillation, the POD modes appear in pairs.  A side-by-side comparison of the dominant POD modes with the first few oscillatory DMD modes reveals a striking semblance of POD modes 1 and 2 with the real and imaginary parts of DMD mode 1.  The same semblance can be found when comparing for POD modes 3 and 4 with the real/imaginary parts of DMD mode 2, and so on for the higher order modes.  Indeed, one can quantify the similarity between these mode shapes by taking an inner product between the DMD mode and a complex vector whose real and imaginary components are formed by the associated POD mode.  Performing such an analysis confirms the strong similarity between POD modes and DMD modes for the periodic cylinder wake [\citen{chen2011variants}].  Having established this similarity, it now becomes evident that one can plot the magnitude and phase of pairs of POD modes, much in the way that we had done for the DMD modes.  Doing so can provide additional interpretable insights into the dynamic significance of these modal structures.

We note that spatial modes from POD and DMD are not identical to each other, in general.  However, they are identical when the flow is periodic in time, as is the case for the cylinder wake at $Re = 100$ considered here.  We further note that DMD modes are not necessarily orthogonal to each other, whereas POD modes are.  
This is an important point to remember when spatial modes are used to form a basis set for use in, e.g.,~reduced-order modeling to be discussed later in Section~II.D.1.

Although POD and DMD yield the same spatial modes in the context of the periodic cylinder wake, DMD offers additional information that POD is not equipped to provide.  Indeed, each dynamic mode consists of spatial information that is embedded in each DMD mode (see \fig \ref{fig:dmd_cylinder_compiled}(a)) \emph{and} temporal information that is embedded in each DMD eigenvalue (see \fig \ref{fig:dmd_cylinder_compiled}(b)).  Since dynamic modes correspond to the spectral decomposition of a best-fit linear operator that maps one snapshot to the next, the dynamics of each DMD mode is comprised of only a single oscillation frequency and a growth/decay rate.  To determine the oscillation frequency $f_i$ and growth/decay rate $g_i$ of spatial mode $i$, we can make use of the associated DMD eigenvalue $\lambda_i$ through $f_i = \angle \lambda_i/(2\pi \delta t)$ and $g_i = \log|\lambda_i|/\delta t$ with $\delta t$ denoting the uniform sampling increment between snapshot pairs.

In addition to spatial structures (DMD modes) and their simple temporal characteristics (DMD eigenvalues),
one can also use the so-called DMD amplitudes to determine the relative contribution of each mode to a particular realization of the system.  We note here that different definitions for DMD amplitudes are used throughout the literature.  However, two common definitions are (1)~DMD amplitudes calculated solely from the initial snapshot, which requires the reciprocal DMD modes, and (2)~DMD amplitudes based on all snapshots, which requires a Vandermonde matrix constructed from the DMD eigenvalues.  For simplicity, we make use of the former definition in \fig \ref{fig:dmd_cylinder_compiled}(b).

A key lesson to take away from the analysis of a cylinder wake is that the data collection step is an important consideration.  Although one may have access to simulation or experimental data from wake start-up through the periodic evolution on the limit cycle, one rarely benefits from ``throwing'' DMD or another modal analysis technique at this full dataset when particular physical questions are of interest.  We can consider the study by Chen and Rowley~[\citen{chen2011variants}], in which the dynamics of a cylinder wake are first split into three distinct regimes: near-equilibrium linear dynamics, post-Hopf bifurcation transient dynamics, and periodic limit cycle dynamics.  In performing their analysis, care is taken to consider only snapshots from each of these three regimes independently.  Not doing so would contaminate the results of the modal analysis, and would arguably lose the interpretability that was being sought.  A similar procedure is performed by Bagheri~[\citen{Bagheri:2013}].  In that work too, the evolution of the cylinder wake is divided into four intervals of interest, each associated with a different time scale.  Although the cylinder wake is a commonly studied and fairly well-understood flow, both of these works demonstrate how supplementary signals, such as force response data (i.e.,~lift in~[\citen{chen2011variants}] and drag in~[\citen{Bagheri:2013}] can be used to delineate between intervals of flow evolution that should be treated separately.  By collecting snapshots from these distinct intervals, we can be better equipped to uncover dynamically relevant features and to better understand the dynamical processes underlying the fluid flow.  Indeed, in our analyses of the cylinder wake above, we ensured that data was only collected once the wake had reached the limit cycle state.

Finally, it is important to highlight a practical caution for users who are interested in conducting DMD analysis of experimental datasets.
DMD has been observed to exhibit sensitivities~[\citen{duke2012error}] and shown to yield biased results~[\citen{hemati2015tls,dawson2016dmdnoise}] when the snapshot data under consideration possess measurement uncertainties, e.g., due to sensor noise.
Thus, users are strongly encouraged to consider noise-robust variants of DMD when analyzing experimentally acquired datasets~[\citen{hemati2015tls,dawson2016dmdnoise,hematistdmd2016}].

\subsection{Linear global stability analysis}

In the above discussions, we used the data-based techniques to study the modal structures generated in the wake of a cylinder.  In this section, we will obtain modal structures directly from the linear evolution operator of the Navier--Stokes equations.  To examine the cause of unsteadiness in the flow, we resort to stability analysis.  There are two types of stability analysis that can reveal the characteristics of flow instabilities; namely, local and global stability analyses.  We focus here on the latter approach, which is suitable for examining instabilities that have global coupling and coverage over the domain of interest.  The main difference between global stability analysis [\citen{Theofilis:PAS03,Theofilis:ARFM11}] and the data-based modal analysis techniques discussed above (POD, DMD) is that the global stability analysis requires access to the base flow and the linearized Navier--Stokes operator based on a numerical discretization (e.g., finite difference, finite volume, or spectral method code), while the data-based methods do not.  

For global stability analysis, we need the base flow $\overline{\boldsymbol{q}}$ about which to linearize the Navier--Stokes equations.  The base state $\overline{\boldsymbol{q}}$ should be the stable or unstable steady state solution (equilibrium state).  See the first overview paper on how to find these states [\citen{Taira_etal:AIAAJ17}].  By decomposing the state variable $\boldsymbol{q}$ into the base state and a perturbation $ \boldsymbol{q}'$ (i.e., $\boldsymbol{q} = \overline{\boldsymbol{q}} + \boldsymbol{q}'$), we arrive at the linearized Navier--Stokes equations in discrete form:
\begin{equation}
    \frac{{\mathrm d}\boldsymbol{q}'}{{\mathrm d}t} = L_{\overline{\boldsymbol{q}}} \boldsymbol{q}',
    \label{eq:linear_dynamics}
\end{equation}
where the linear operator $L_{\overline{\boldsymbol{q}}}$ is dependent on the base state $\overline{\boldsymbol{q}}$.  By expressing the small perturbation $|\boldsymbol{q}|' \ll |\overline{\boldsymbol{q}}|$ as
$\boldsymbol{q}'(\boldsymbol{x},t) = \hat{\boldsymbol{q}}(\boldsymbol{x}) \exp(i \omega t)$,
we arrive at 
\begin{equation}
    L_{\overline{\boldsymbol{q}}}  \hat{\boldsymbol{q}} = i \omega \hat{\boldsymbol{q}}.
\end{equation}
This equation casts the stability analysis of the flow field in terms of an eigenvalue problem.  The eigenvalue $i \omega$ reveals the growth/decay rate $Im(\omega)$ and frequency $Re(\omega)$ of each spatial eigenvector $\hat{\boldsymbol{q}}$ that is found from this analysis.  Thus, we can solve this eigenvalue problem for the dominant eigenvalues and the corresponding eigenvectors to determine the spatial profiles of the instabilities.  Alternatively, one can also time integrate Eq.~(\ref{eq:linear_dynamics}) to determine the dominant mode, as an initial value problem.  Note also that the construction of the eigenvalue problem for linear stability analysis used here is standard, but leads to a reversal of roles in the real and imaginary components of eigenvalues when compared to the DMD analysis.  While it is possible to consider the use of a time-average (or ensemble-average) state as the base flow, linear stability analysis would not hold, since such state in general is not an equilibrium state.  However, the use of a time-average base flow may provide some insights as a model and is used as a precursor to examine the stability property for its resolvent analysis [\citen{Yeh:JFM2019}] (see other flow examples below).

Now, let us consider the application of linear global stability analysis to cylinder wakes.  One of the important insights that can be gained from this analysis is the onset of the wake instability, i.e., the von K\'arm\'an shedding, which appears at a critical Reynolds number of $Re_\text{crit} \approx 46$.  The onset of instability can be identified when the eigenvalues from stability analysis cross over from the stable to the unstable complex plane, as the Reynolds number $Re$ increases beyond its critical value $Re_\text{crit}$.  The appearance of this type of instability is called the Hopf bifurcation.  We compile the critical Reynolds numbers where the flow is observed to initiate the von K\'arm\'an shedding in Table \ref{table:cylinder_re}.  Included are the critical Reynolds numbers found from careful experiments performed by Taneda [\citen{Taneda:JPSJ56}] and Strykowski and Sreenivasan [\citen{Strykowski:JFM90}].  On the numerical side of the studies, the global stability analysis performed by Zebib [\citen{Zebib:JEB87}] and Jackson [\citen{Jackson:JFM87}] have predicted the critical transitions well.  Here, we take a broad definition of global stability analysis, especially for papers from earlier stability studies when computational resources were limited.  In most of the earlier studies, the modes are not reported but the eigenvalues are reported in detail.  More recently, the stability modes of the cylinder wake at $Re = 50$ have been examined by Abdessemed et al.~[\citen{Abdessemed:PF09}] in a broader context.  What is striking from their analysis is the resemblance of the modal shapes between the energetic mode (such as the dominant POD mode shown above) and the stability mode.  This observation suggests that the instability in the flow causes the wake to oscillate, resulting in the emergence of the von K\'arm\'an shedding in the full nonlinear flow.  Since the Reynolds number considered in this case is near the bifurcation point of $Re_\text{crit}$, the mode shapes of the dominant instability and that of POD analysis are expected to be similar [\citen{sipp2007global}].

\begin{table}[t]
\small
    \centering
    \begin{tabular}{lllll} \hline
        Transition                  & References                        & $Re_\text{crit}$  & $St_\text{crit}$  & Analysis \\ \hline\hline
        von K\'arm\'an shedding (2D)    
                                    & Taneda [\citen{Taneda:JPSJ56}]    & 45                & -                 & Experimental \\ 
                                    & Provansal, Mathis, \& Boyer [\citen{Provansal:JFM87}]      
                                                                        & 47                & 0.12              & Experimental \\
                                    & Strykowski \& Sreenivasan [\citen{Strykowski:JFM90}]       
                                                                        & 46                & 0.12              & Exp/Num \\
                                    & Zebib [\citen{Zebib:JEB87}]       & 45                & 0.11-0.13         & Stability \\
                                    & Jackson [\citen{Jackson:JFM87}]   & 46.184            & 0.138             & Stability \\
       Mode A (3D)                 & Williamson [\citen{Williamson:PF88b, Williamson:JFM96}]                       
                                                                        & 170-180           & -                 & Experimental \\ 
                                    & Barkley \& Henderson [\citen{Barkley:JFM1996}]
                                                                        & 189               & -                 & Stability \\
        Mode B (3D)                 & Williamson [\citen{Williamson:PF88b, Williamson:JFM96}]
                                                                        & 230-260           & -                 & Experimental \\ 
                                    & Barkley \& Henderson [\citen{Barkley:JFM1996}]
                                                                        & 259               & -                 & Stability \\
        \hline
    \end{tabular}
     \caption{Compilation of transition Reynolds numbers determined from experiments and stability analyses.}
    \label{table:cylinder_re}
\end{table}

The linear global stability analysis can be further extended to periodic base states through Floquet analysis [\citen{Taira_etal:AIAAJ17}].  Through such extension, we can determine the appearance of three-dimensional instabilities known as modes A and B, which appear at $Re_A \approx 189$ and $Re_B \approx 259$, respectively [\citen{Williamson:PF88b, Williamson:JFM96}].  Barkley and Henderson [\citen{Barkley:JFM1996}] and Abdessemed et al.~[\citen{Abdessemed:PF09}] have examined such transitions carefully using stability analysis.  We append the theoretical prediction of eigenvalues from the stability analysis in Table \ref{table:cylinder_re}.  The visualizations of the three-dimensional A and B modes, while not shown here, can be found in [\citen{Barkley:JFM1996}], which agree well with the dye visualizations presented in [\citen{Williamson:JFM96}].  With the identification of the three-dimensional stability modes, we can observe the regions from which three-dimensional instabilities are given birth.  We also note in passing that we can further consider modal structures responsible for transient growth, as reported by Abdessemed et al.~[\citen{Abdessemed:PF09}].

\subsection{Flow Modeling}
\label{sec:cylinder_modeling}

\subsubsection{Galerkin modeling}
\label{sec:galerkin_modeling}

We have now seen that the POD modes determined from the snapshots can represent the flow field accurately with remarkable reduction in dimensionality.  Instead of requiring a large number of grid points to represent the flow ($n = \mathcal{O}(10^5-10^6)$), we can simply reconstruct the flow field using a small set of POD modes.  In the case of laminar cylinder flow, 8 modes can capture the unsteadiness very well, as discussed above.  Using the low-dimensional representation of the flow field with POD modes, we can model the dynamics of the flow field.  Here, we present this reduced-order modeling technique based on the Galerkin projection approach [\citen{Holmes96,Aubry:JFM88,Noack11}].  This approach can provide a small set of ordinary differential equations in terms of the POD coefficients ${\boldsymbol{a}} \in \mathbb{C}^r$, where $r \ll n$, to describe the dynamics of the flow.

Let us consider the velocity field to be expressed as a superposition of the POD modes
\begin{equation}
    \boldsymbol{u}(t,\boldsymbol{x}) = \sum_{j=0}^r a_j(t) \boldsymbol{\varphi}_j(\boldsymbol{x}),
    \label{eq:PODexpansion}
\end{equation}
where $a_0 = 1$ and $\boldsymbol{\varphi}_0$ represents the mean field.
We substitute this series into the incompressible Navier--Stokes equations and find that 
\begin{equation}
    \frac{\partial}{\partial t} \sum_{j=0}^r a_j(t) \boldsymbol{\varphi}_j(\boldsymbol{x})
    +  \sum_{j=0}^r a_j(t) \boldsymbol{\varphi}_j(\boldsymbol{x}) \cdot \nabla
    \sum_{k=0}^r a_k(t) \boldsymbol{\varphi}_k(\boldsymbol{x})
    = - \nabla p + \frac{1}{Re} \nabla^2  \sum_{j=0}^r a_j(t) \boldsymbol{\varphi}_j(\boldsymbol{x})
\end{equation}
\begin{equation}
    \nabla \cdot \sum_{j=0}^r a_j(t) \boldsymbol{\varphi}_j(\boldsymbol{x}) = 0.
\end{equation}
The second equation is the continuity equation, which is automatically satisfied by each and every POD mode.  Hence, we only need to consider the momentum equation.  To project these dynamics onto the POD modes, we take an inner product of the above equation with $\boldsymbol{\varphi}_i(\boldsymbol{x})$ which yields
\begin{equation}\label{eq:Galerkinproj}
    \frac{{\mathrm d} a_i}{{\mathrm d}t} = 
    \sum_{j=0}^r F_{ij} a_j
    + \sum_{j=0}^r \sum_{k=0}^r G_{ijk} a_j a_k ,
\end{equation}
where 
$F_{ij} = - Re^{-1} \langle \boldsymbol{\varphi}_i, \nabla^2 \boldsymbol{\varphi}_j \rangle$ 
and 
$G_{ijk} = - \langle \boldsymbol{\varphi}_i, \boldsymbol{\varphi}_j \cdot \nabla \boldsymbol{\varphi}_k \rangle$ 
for $i = 1, \dots, r$. To arrive at the above model, we used the orthogonality property of POD modes (i.e., $\langle \boldsymbol{\varphi}_i, \boldsymbol{\varphi}_j \rangle = \delta_{ij}$).  
From the initial condition for the temporal coefficients $a_i(t_0) = \langle \boldsymbol{u}(t_0,\boldsymbol{x}), \boldsymbol{\varphi}_i(\boldsymbol{x})\rangle$, we can now simply integrate these ODEs to predict the dynamics of the modes.  To reconstruct the full flow field, we simply use \eq (\ref{eq:PODexpansion}).  The pressure gradient term drops out from the Galerkin model due to the boundary condition for most flows.  Extensive discussions on the treatment of pressure term and its influence on the model are provided in Holmes [\citen{Holmes96}] and Noack et al.~[\citen{Noack:JFM03}].  Modeling efforts by incorporating the pressure POD modes have also been considered to account for the pressure effects~[\citen{Bergmann:JCP09}].

For modeling the cylinder wake, Deane et al.~[\citen{deane1991galerkin}] noticed that the use of only 4 POD modes results in slow growth of the wake oscillation amplitudes without bounds.  However, the reduced-order model (ROM) with 6 POD modes improved the accuracy of the model.  Deane et al.~made a couple of important observations.  First is the validity of the ROM modes; for a fixed Reynolds number, the model predicts the behavior of the flow well.  However, the use of the mean field and the POD modes from one Reynolds number does not appear to accurately extend to other Reynolds number.  Second observation is the accuracy of the model over a long time frame.  These POD-based reduced-order model that predicts the dynamics well for a short duration can deviate over a long time.  

\begin{figure}
    \centering
    \includegraphics[width=0.35\textwidth]{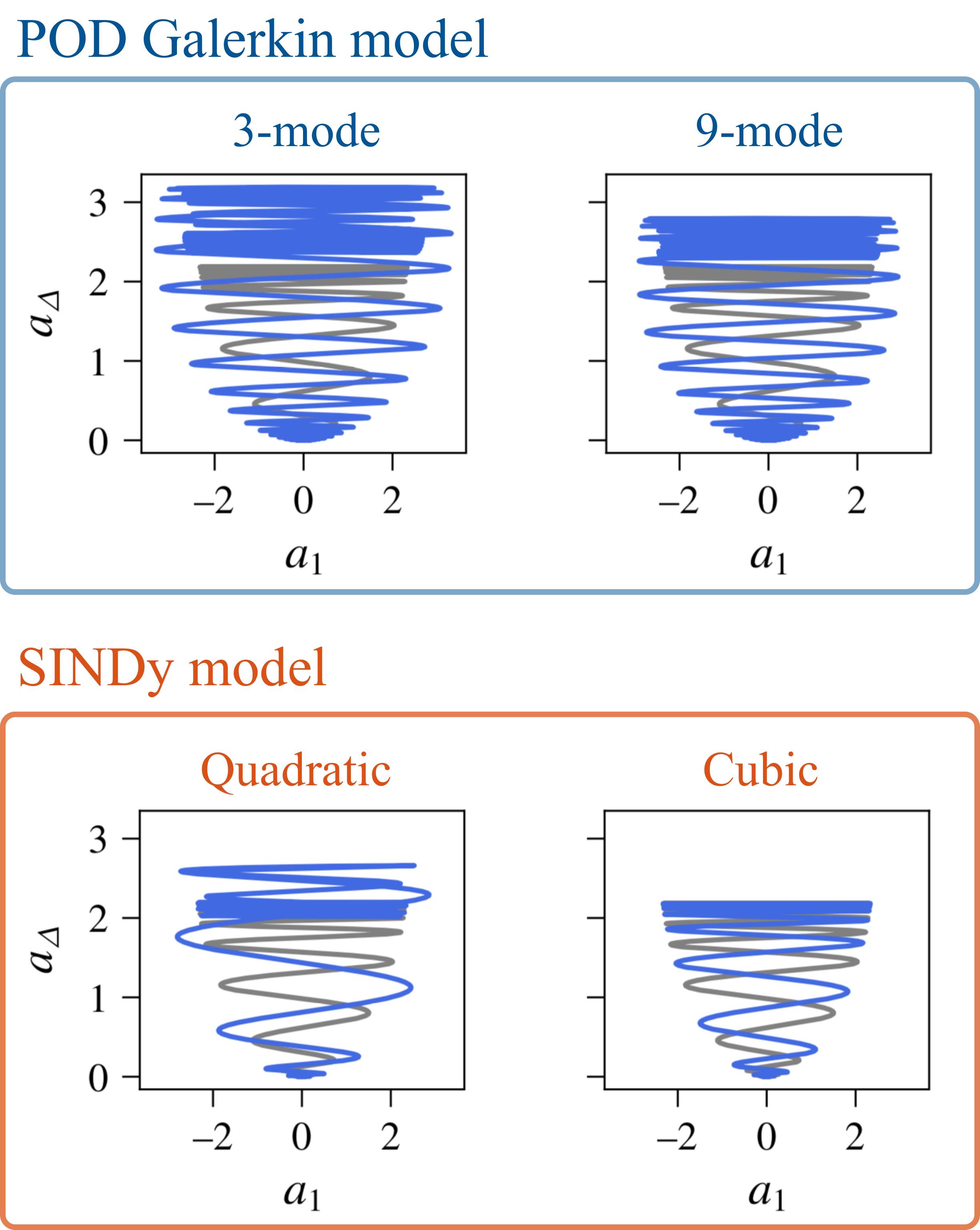}
    \caption{Galerkin projection and SINDy models for cylinder flow at $Re = 100$.  The coefficients capture the transient dynamics of cylinder flow developing the von K\'arm\'an shedding instability from the unstable steady state at the bottom of the paraboloid [\citen{Loiseau2017jfm}].  Shown are reference (gray) and model (blue) trajectories.  Reprinted with permission from Cambridge University Press.
    }
    \label{fig:cylinder_model}
\end{figure}

Noack et al.~[\citen{Noack:JFM03}] noted the importance of the base flow and constructed a POD-based reduced-order model that can capture transients from an unstable equilibrium to an asymptotic shedding state (limit cycle oscillation).  In addition to the POD modes obtained from the flow field data, they supplemented the set of POD basis with the shift mode $\boldsymbol{\varphi}_\Delta$.  This additional shift mode amounts to the difference between the mean flow and the equilibrium state, with the modal components projected out, such that 
\begin{equation}
    \boldsymbol{\varphi}_\Delta = \frac{\boldsymbol{\varphi}_\Delta^b}{\| \boldsymbol{\varphi}_\Delta^b \|}, 
    \quad \text{where} \quad
    \boldsymbol{\varphi}_\Delta^b 
    = \boldsymbol{\varphi}_\Delta^a 
    - \sum_{i=1}^r \left< \boldsymbol{\varphi}_\Delta^a, \boldsymbol{\varphi}_i \right> \boldsymbol{\varphi}_i
    \quad \text{and} \quad
    \boldsymbol{\varphi}_\Delta^a 
    = \boldsymbol{u}_0 - \boldsymbol{u}_s.
\end{equation}
Here, ${\boldsymbol{u}}_0$ is the time-averaged flow and ${\boldsymbol{u}}_s$ is the steady-state solution to the Navier--Stokes equations.  By adding this shift mode to the set of basis functions, the Galerkin-based reduced order model can be improved to model the transient dynamics well.  Such prediction is generally difficult without the shift mode.  This model with the shift mode can capture the transient effects with only 3 modes (shift, first, and second modes), which is a significant reduction in the dimension to describe the emergence of wake instability.  Shown in \fig \ref{fig:cylinder_model} are the POD-based Galerkin model results compared with the full DNS.  Shown on the left are the 3 and 9-mode results using the approach of Noack et al.~[\citen{Noack:JFM03}].

The present discussion on the Galerkin projection model was based on the use of orthogonal POD modes.  However, a set of non-orthogonal modes, such as the DMD modes, can be used instead.  The resulting model for non-orthogonal modes would include additional terms, yielding a Petrov-Galerkin model.  It is also possible to introduce adjoint modes to use bi-orthogonality to derive a reduced-order model [\citen{Zhang:AIAAAviation17}].  Another approach is to incorporate concepts from network science and machine learning to construct a modal-network model [\citen{Nair:PRE18}].

Despite the success of Galerkin projection, there are a number of practical issues that arise when using it to develop reduced-order models.  
First, the expansion in Eq.~\eqref{eq:PODexpansion} is typically truncated at a low order $r\ll n$ in order to develop an efficient model in terms of a few dominant coherent structures.  
However, truncating the modal expansion removes the effects of low-energy modes, which might be dynamically important. 
When these terms are removed, subtle imbalances appear in the quadratic nonlinearities, which often eventually lead to long-term instability.  
It is possible to correct for these issues by enforcing energy-preserving symmetries in the quadratic terms in the Galerkin projection process~[\citen{Noack11,balajewicz2013jfm,Carlberg2015siamjsc,Schlegel2015jfm}].
It may also be the case that for complex, multiscale dynamics, the flow structures are not adequately captured in a low-dimensional linear subspace; because Galerkin projection models typically scale with $\mathcal{O}(r^3)$, these models quickly become more expensive than the original full-order model without further measures.  
Finally, POD modes generally deform with changing flow conditions and geometries, so that a given model derived at one flow condition may have restricted utility for other conditions.  
A number of these issues will be discussed in the next section, and also in the outlook in section~\ref{sec:outlook}.

\subsubsection{Sparse identification of nonlinear dynamics}

As an alternative to Galerkin projection of the governing equations onto an orthogonal POD basis, it is possible to identify a nonlinear dynamical system in this subspace by data-driven regression.  In particular, the sparse identification of nonlinear dynamics (SINDy) algorithm~[\citen{Brunton2016pnas}] may be used to discover a low-order model based on time-series data of the POD coefficients as the system evolves.  Given a vector of POD coefficients, ${\boldsymbol{a}}$, it is possible to represent the right hand side of the dynamics of ${\boldsymbol{a}}$ as a linear combination of basis functions $\theta_j({\boldsymbol{a}})$ in a library:
\begin{equation}
    \frac{{\rm d} {\boldsymbol{a}}}{{\rm d}t} 
    = {\boldsymbol{f}}({\boldsymbol{a}}) 
    \approx \sum_{j=1}^p \theta_j({\boldsymbol{a}}) \xi_j.
\end{equation}
The SINDy algorithm seeks a sparse vector of coefficients ${\boldsymbol{\xi}}$, indicating that as few terms $\theta_j({\boldsymbol{a}})$ are active in the dynamics as possible.  This is achieved via modern methods in sparse regression, and helps to ensure that the resulting models are both interpretable.  This approach was recently extended to model fluid systems by Loiseau \etal~[\citen{Loiseau2017jfm,Loiseau2018jfm}].  It was shown that known constraints can be incorporated in the SINDy regression framework, such as energy conserving constraints on the quadratic nonlinearities for incompressible flows [\citen{Loiseau2017jfm}]. In particular, it is known that a particular skew-symmetry in the quadratic nonlinearities gives rise to energy conservation in incompressible flows, and it is possible to enforce this model structure in the sparse regression procedure via Lagrange multipliers.  In general, there is a growing effort, especially in fluid mechanics, to incorporate known symmetries, constraints, and conservations laws into various machine learning algorithms [\citen{Brunton2019arfm}]. 

Furthermore, unlike in Galerkin projection, where the nonlinear terms in the reduced-order model reflect those in the governing equations, in SINDy, it is possible to include higher-order nonlinearities, which may serve to account for the effects of truncated terms in the POD expansion.  A comparison of SINDy models and standard Galerkin projection is shown in \fig \ref{fig:cylinder_model}, where it can be seen that a SINDy model with cubic nonlinearities nearly perfectly captures the true dynamics. This approach was later shown to be effective on sensor-based coordinates, such as lift and drag measurements, removing the need for full-state data and POD analysis, and bypassing the mode deformation associated with changing flow conditions.  These reduced-order models can be used to stabilize the wake shedding for drag reduction [\citen{Brunton2015amr}].


\section{Wall-bounded flows}
\label{sec:BL}

Wall-bounded flows are one of the most ubiquitous flows that arise in the study of fluids systems.  Wall-bounded shear flows have important differences from those considered in Section \ref{sec:cylinder}, which may exhibit oscillator-type dynamics characterized by a single dominant frequency and length scale. In contrast, wall-bounded flows at sufficiently large Reynolds number can exhibit energetic structures at a broad range of length- and time-scales.  Wall-bounded configurations including channel, pipe, Couette, and boundary layer flows share similarities both in terms of flow physics and analysis methods.  Here, the focus will be on flow through a uniform channel with infinite extent in the streamwise and spanwise directions.  The assumption of spatial homogeneity assumed here is common in modal analysis, as spatial Fourier modes can be used to represent these spatially homogenous directions [\citen{Holmes96}].  However, this simplification can come at the cost of inefficiencies for identifying and modeling localized or spatiotemporally developing structures.  Nonetheless, we adopt this simplification, as the assumption of spatial homogeneity typically reduces the amount of data required for data-driven methods, and reduces the computational requirements for operator-based decompositions.

Famously, laminar flow in a channel becomes linearly unstable at a Reynolds number (based on channel half-height) of $Re=5\,772$, as first computed precisely by Orszag [\citen{orszag1971os}].  However, transition to turbulence may be triggered and sustained at a much lower Reynolds number than predicted by this linear ({\it{modal}}) stability analysis.  Indeed, the linear dynamics predict that perturbations about the laminar equilibrium state can exhibit significant growth prior to subsequent decay---a phenomenon known as transient energy growth, which can be studied using {\it non-modal} stability analysis~[\citen{Schmid:ARFM07}].  Transient energy growth is commonly attributed to the high-degree of non-normality of the linearized Navier-Stokes operator, which is observed in numerous wall-bounded flows.  The purpose of this section is to demonstrate how modal decomposition techniques may be applied to this class of flows, and to summarize and compare typical results from each method. We will not attempt to provide a comprehensive analysis of every aspect of the flow physics, nor a complete summary of the substantial body of prior work using modal decomposition techniques on this class of problems.  In this section, we will first consider in detail the  case of a stable linearized channel flow in Section III.\ref{sec:linchan}, before discussing how similar methods may be extended to study turbulent (Section III.\ref{sec:turbchan}) and spatially developing (Section III.\ref{sec:bl}) wall-bounded flows.

\subsection{Linearly stable laminar channel flow}
\label{sec:linchan}

\begin{figure}
\centering
   \includegraphics[width=0.45\textwidth]{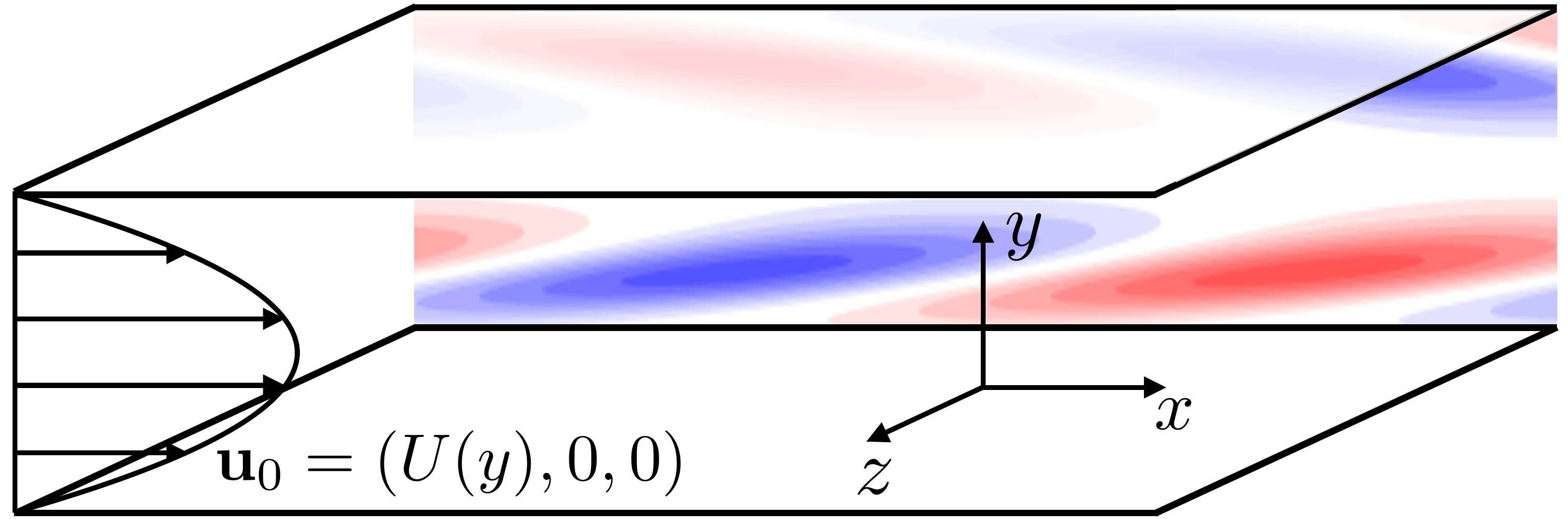}
   \caption{The channel geometry superposed with the mean profile and a 2D slice of the leading resolvent response mode for this system at $Re = 2\,000$, $k_x = 0.25$ $k_z = 2$, and $c_r = 0.578$.}
   \label{fig:channel_diagram}
\end{figure}  

We consider a system at a half-height based Reynolds number of $Re=2\,000$.  This section will focus only on the dynamics with streamwise and spanwise wavenumbers of $k_x = 0.25$ and $k_z = 2$.  The dynamics are linearized about a laminar equilibrium state (i.e.,~a  parabolic streamwise velocity profile).  This simple example highlights certain features of wall-bounded flows, and serves as a testbed to compare many of the modal decomposition methods considered in this paper.  There are many references that give a more comprehensive treatment of this system [\citen{Schmid01,Jovanovic05}] with some of the analysis presented here similar to that described in [\citen{Kim:ARFM07, Schmid01, Schmid:ARFM07, Rowley-ijbc05, Rowley:ARFM17}].  For this example, we formulate the problem in terms of the Orr-Sommerfeld and Squire equations and show results using the wall-normal velocity and vorticity.
 
In contrast to the approach taken in Section \ref{sec:cylinder}, we first consider operator-based analysis of this system, before progressing to data-driven analysis (where the choice of data will be informed by the results of the operator-driven analysis). 
To start with, we consider the stability properties of the linear operator for this system. In addition to studying the asymptotic stability (governed by the eigenvalues, or spectrum of the operator), we also consider the pseudospectrum, which can be viewed as a measure of how close a given point in the complex plane is to being an eigenvalue. More formally, the $\epsilon$-pseudospectrum of a linear operator $L$ is the set of values $z$ in the complex plane satisfying the following equivalent conditions for a given $\epsilon$:
\begin{enumerate}
    \item $\|L\boldsymbol{q}  - z \boldsymbol{q}\| \leq \epsilon$ for some ``pseudoeigenvector" $\boldsymbol{q}$
    \item $(L+E) \boldsymbol{q} = z \boldsymbol{q}$ for some perturbation operator $E$, with $\|E\|\leq \epsilon$
    \item $\|(zI - L)^{-1}\| \geq \epsilon^{-1}$
\end{enumerate}
The operator $(zI - L)^{-1}$ is the resolvent operator associated with $L$ for a given $z$, with the associated resolvent norm $\|(zI - L)^{-1}\|$. 
Detailed discussion of pseudospectral theory in the context of fluid flows are offered in [\citen{ReddySchmidHenningson,Schmid:ARFM07}].  

\begin{figure}
\centering
\small
    \begin{tabular}{ll}
    (a) & (b) \\
    \includegraphics[width=0.48\textwidth]{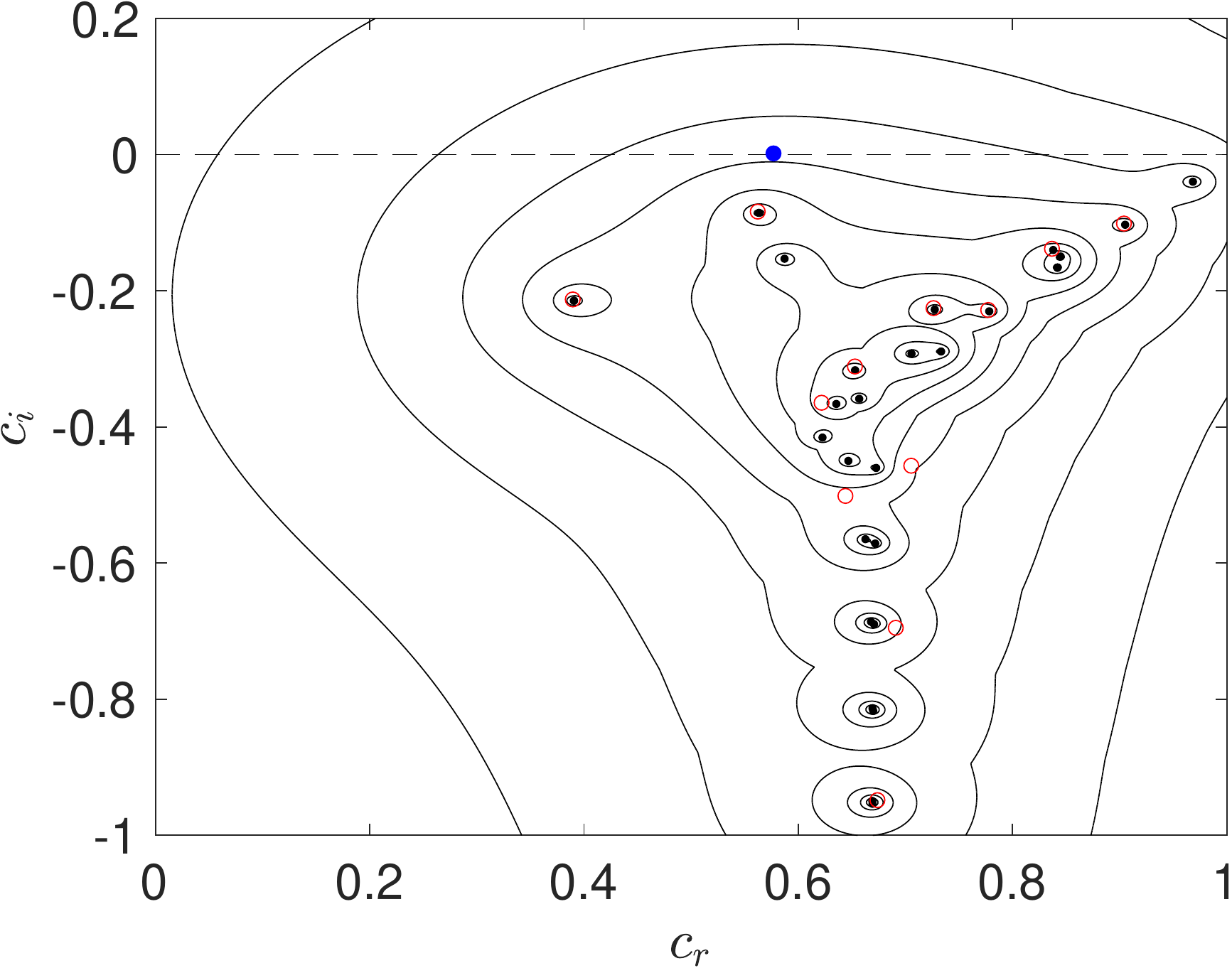} &
    \includegraphics[width=0.45\textwidth]{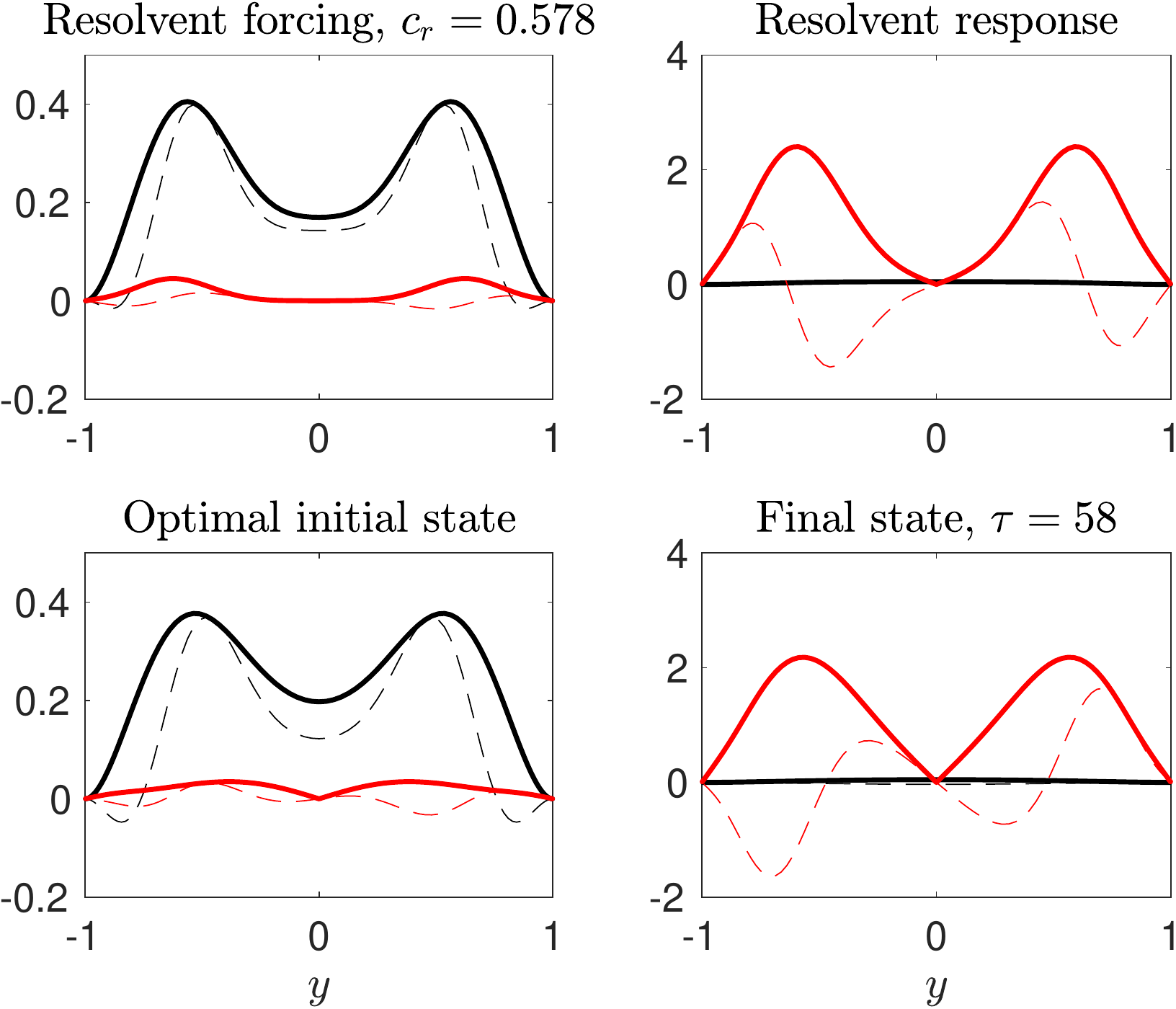}
    \end{tabular}
    \caption{(a) Eigenvalues (expressed as a complex growth rate $c = \lambda/k_x $) and contours of the pseudospectra for channel flow with $Re = 2000$, $k_x = 0.25$ and $k_z = 2$. DMD eigenvalues identified for a trajectory giving optimal transient growth are shown with unfilled circles, while the filled circle at $c_r = 0.578$ represents the real frequency corresponding to maximum resolvent norm. (b) Leading resolvent forcing and response modes corresponding to a wavespeed ($c_r = 0.578$) leading to maximum amplification, and initial and final (maximally amplified) states along a trajectory leading to maximal energy growth. All modes are scaled to be of unit norm, with black and red lines corresponding to wall-normal velocity and vorticity fields, and solid and dashed lines representing the magnitude and real component of these fields.
    }
    \label{fig:channel_spectrum}
\end{figure}

The eigenvalue spectrum for this stable system, along with contours of the pseudospectra for various values of $\epsilon$, are shown in Fig.~\ref{fig:channel_spectrum}(a).
By Squire's theorem, the system would become less stable if the spanwise wavenumber were reduced to 0.  Even when the system is asymptotically stable, the non-normal nature of the associated linear operator  renders it susceptible to both transient energy growth over finite time-horizons, and high amplification of external disturbances/inputs.  
Note that in the case of a normal operator, the $\epsilon$-pseudospectrum consists of the union of concentric disks of radius $\epsilon$ about each eigenvalue. For non-normal operators such as that considered in Fig.~\ref{fig:channel_spectrum}(a), the $\epsilon$-pseudospectrum can be far larger, and can deform away from being a union of concentric disks. In this case, the union of concentric disks of radius $\epsilon$ about each eigenvalue instead gives a lower bound for the $\epsilon$-pseudospectrum, with an upper bound given by the equivalent union of disks of radius $\kappa\epsilon$, where $\kappa$ is the condition number of the operator.

The maximum amplification of the system to a single-frequency input is given by the norm of the resolvent operator associated with the linear system for that frequency, which is equivalent to the contours of pseudospectra plotted in Fig.~\ref{fig:channel_spectrum}(a).  We are typically interested in purely oscillatory disturbances, which correspond to the dashed line in Fig.~\ref{fig:channel_spectrum}(a). The optimal disturbance and response at a wave speed ($c_r  = \omega/k_x = 0.578$) leading to maximum amplification for this system are shown in Fig.~\ref{fig:channel_spectrum}(b) (which corresponds to the filled circle in Fig.~\ref{fig:channel_spectrum}(a)). A contour plot through a spanwise-constant slice of the domain of the resolvent response mode for these parameters is also shown in Fig.~\ref{fig:channel_diagram}. This optimal disturbance and response may be obtained respectively from the leading right and left singular vectors of the resolvent operator associated with this frequency.

\begin{figure}
\centering
   \includegraphics[width=0.99\textwidth]{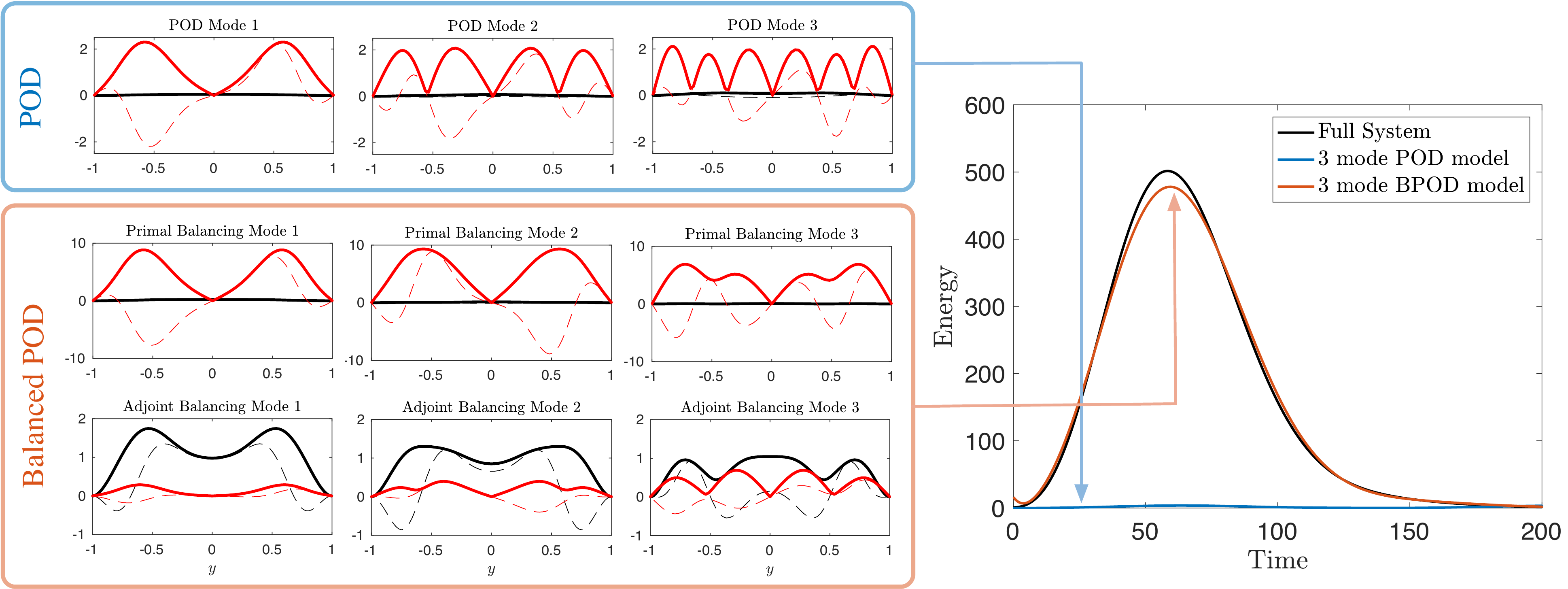}
   \caption{Development of POD and Balance POD based models.  (Left) POD, and balancing primal and adjoint mode amplitudes (solid lines) and real components (dashed) for the wall-normal velocity (black) and vorticity (red).  (Right) Evolution of the energy of the system along a trajectory leading to maximal finite-time energy growth, compared with predictions from the 3-mode POD and Balanced POD models.}
   \label{fig:channel_ROM}
\end{figure}

The initial and final conditions of the trajectory giving maximal energy growth for this system are shown in Fig.~\ref{fig:channel_spectrum}(b).  The maximum energy is attained by the system at a time horizon $\tau = 58$, where the initial and final conditions giving maximum energy growth can be obtained from the leading right and left singular vectors of the finite-time propagation operator $\exp(L\tau)$.  The evolution of the energy of the system for this trajectory will be shown in Fig.~\ref{fig:channel_ROM}. Note that transient growth may be formally related to pseudospectral/resolvent analysis via the Kreiss constant [\citen{Trefethen05}], as discussed in the context of channel flow in [\citen{ReddySchmidHenningson,Schmid:ARFM07}].

Thus far, we have analyzed this system through a study of the operator itself.  We now give attention to data-driven modal decomposition methods.  We will focus on data collected on the trajectory giving the largest energy growth, with an initial condition as shown in Fig.~\ref{fig:channel_spectrum}(b). 
Performing DMD on this trajectory (collected with a time step $\delta t = 0.01$) gives eigenvalues as shown by the open circles in Fig.~\ref{fig:channel_spectrum}(a).  We observe that DMD identifies some, but not all of the eigenvalues of the system. Physically, DMD identifies those modes which are active in the given dataset.  Since DMD is a data-driven method, it cannot take advantage of rescaling and normalization that are typically applied when using iterative methods such as an Arnoldi procedure [\citen{Schmid:JFM10}]. Note in particular that for this system, the eigenvalues near the intersection between the eigenvalue branches are particularly susceptible to perturbation, which is again related to the nonnormality of the system.  Furthermore, while the operator identified using DMD shares only a subset of the true eigenvalues of the full operator, it is able to reconstruct the data that was used for its identification.  This is not surprising since the underlying dynamical system is linear. 

POD modes identified from this dataset are shown in Fig.~\ref{fig:channel_ROM}. We observe that the leading POD modes are dominated by the wall-normal vorticity component, and also that only a few modes are required to account for the vast majority of the energy present in the data. 
We may use these modes as a basis for projection of the governing equations, to obtain a reduced-order model for the system dynamics.  Note that this is the same procedure presented in Section II.D.1, but here we consider the simpler case of a linear system. In particular, if the basis of POD modes to be used for projection are given by the columns of a matrix $\Phi$, then the reduced-order operator $\tilde L$ can be obtained from the full operator $L$ by 
\begin{equation}
\label{eq:PODproj}
    \tilde L = \Phi^*L \Phi,
\end{equation} 
where $\Phi^*$ is the adjoint of $\Phi$. 
Performing such a projection using the first three POD modes gives a model that poorly reconstructs the trajectory of the data, as shown in Fig.~\ref{fig:channel_ROM}.

This can be understood by the fact that the total energy of the system, and thus the leading POD modes, are dominated by the wall-normal vorticity component, yet the initial condition used contains a substantial wall-normal velocity component.  Note in particular that the leading three POD modes account for less than $4\%$ of the energy of the initial condition.  However, if five modes are used instead, this measure exceeds $95\%$, and an accurate reduced-order model can be obtained.  This demonstrates that total energy content is often not the most important factor for identifying a basis for projection of the full governing equations, since there may be features that are dynamically important, even though they are low in energy.

For a linear system with a given set of system inputs and outputs, a projection-based reduced-order model that best preserves the input-output dynamics may be obtained through balanced truncation [\citen{Moore-81}].  A balanced truncation finds a compromise between system observability and controllability, which are analogous to the total energy content and initial conditions for the system considered here (if one considers the initial conditions as the input matrix for a single-input linear state-space system). Rather than using a single set of modes to form a basis for projection as in Eq.~(\ref{eq:PODproj}) we obtain a reduced-order model by finding separate subspaces that define the basis upon which the reduced dynamics evolve (basis vectors for which can be assembled as columns of a matrix $\Phi$), and the direction of the projection onto that subspace (with a basis given by the columns of $\Psi$, which are bi-orthogonal to those of $\Phi$). The reduced linear operator is then related to the full system by
\begin{equation}
\label{eq:balproj}
    \tilde L =  \Psi^* L \Phi,
\end{equation}
with system inputs and outputs appropriately projected. We refer to the columns of $\Phi$ and $\Psi$ as the primal and adjoint modes, respectively.

While it is feasible to perform balanced truncation directly for the simple system considered here, it can become difficult or infeasible for large systems. Balanced POD [\citen{Rowley-ijbc05}] gives a means to perform approximate balanced truncation from impulse response data from the forward and adjoint systems (which have state propagation operators are given respectively by $L$ and $L^*$). As seen in Fig.~\ref{fig:channel_ROM}, applying this method gives a three-mode model that accurately captures the trajectory of the data, unlike the POD model. For the balanced POD model, we apply output projection [\citen{Rowley-ijbc05,Ahuja:JFM10}] to reduce the dimension of the full-state output down to just seven variables. While the POD model projects the governing equation onto a subspace along a direction orthogonal to the subspace, balanced POD computes a direction of projection which best preserves the dynamics of the system.  We show in Fig.~\ref{fig:channel_ROM} the primal modes onto which the dynamics are projected, and the adjoint modes defining the direction of projection. Note in particular that, while the primal modes are dominated by the wall-normal vorticity (as are the POD modes), the adjoint modes have a substantial contribution from the wall-normal velocity. This allows for a projection that retains sufficient dynamically-important content of the full-system to give an accurate reduced-order model. 

The possibility of capturing input-output dynamics using modal analysis techniques has made balanced truncation and BPOD attractive for active flow control synthesis.  Indeed, modal analysis can guide the choice of actuation and sensing and can also inform designs of open-loop control strategies~[\citen{bhattacharjeeAIAA2018,yaoAIAA2019}].  Further, modal analysis techniques can be tailored to and leveraged for feedback flow control synthesis.  Within the context of channel flow control, numerous model reduction strategies have been developed around modal decomposition techniques, e.g., global mode truncation~[\citen{martinelliPOF2011}] and input-output modeling [\citen{ilakAIAA2008,kalurAIAA2018,kalur2019}].  Recent efforts have demonstrated that a model reduction approach needs to be selected and tailored carefully with respect to the control objective~[\citen{kalurAIAA2018,kalur2019}].  For example, within the context of transient energy growth reduction, the performance of feedback controllers designed on reduced-order models~(ROMs) 
can be quite sensitive to the parameters used in generating the underlying ROMs~[\citen{kalurAIAA2018,kalur2019}].

Interestingly, the physical and descriptive insights offered by modal analysis can potentially guide feedback flow control designs as well, motivating a new perspective for dynamic mode shaping control synthesis, in which the closed-loop spectral properties of a system can be prescribed by appropriate control action [\citen{hematiAIAA2017}].  These same efforts on dynamic mode shaping have recently been used to uncover fundamental performance limitations in commonly employed sensor-based output feedback controllers that are commonly employed in flow control applications.  Indeed, it can be shown that observer-based feedback strategies---in which the flow state is reconstructed from measured sensor outputs, then leveraged for feedback control---can never fully suppress transient energy growth within the context of linearized flows that exhibit transient energy growth in the first place [\citen{hematiAIAAJ2018}].  Furthermore, such strategies have been found to dramatically degrade performance in terms of the worst-case transient energy growth in the linearized channel flow system [\citen{yaoAIAA2018, yaoAIAA2019}].  Alternative output feedback control strategies have been found to be superior in terms of the worst-case transient energy growth performance.  However, most synthesis algorithms suffer from the curse of dimensionality, necessitating the use of reduced-order models to make synthesis tractable in flow control applications.  Control-oriented model reduction based on modal decompositions and other systems-theoretic techniques (e.g., robust/$\mathcal{H}_\infty$ modeling [\citen{jonesJFM2015}]) will play an important role in overcoming these hurdles into the future.

\subsection{Turbulent wall-bounded flows}
\label{sec:turbchan}

This section will discuss how a number of ideas and methods discussed in Section III.\ref{sec:linchan} have been applied to turbulent flows, where spatial homogeneity is assumed in the streamwise and spanwise directions.  Operator-based linear analysis of turbulent flows typically consider mean-linearized governing equations.  While such analyses are generally not able to predict the exact evolution of trajectories as in the case of linear flow, substantial insight into features of turbulent flows can still be gained from consideration of linear operators.  Modal linear stability analysis about wall-bounded turbulent mean flows often gives stable eigenvalues [\citen{reynolds1967stability, delAlamo2006linear}] (though this is not the case for some of the geometries considered in Sections IV and V).  Indeed, the discrepancy between Squire's theorem, whereby the least stable modes are spanwise-constant, and observations of the three-dimensionality of both transition mechanisms [\citen{klebanoff1962}] and structures in fully developed turbulence show the limitations of linear eigenmodes.  The combination of a nonnormal system with nonlinear terms of substantial size means that nonmodal properties make considerations of non-normality particularly important for wall-bounded (or more generally, shear-driven) turbulent flows [\citen{Schmid:ARFM07}].  Operator-based analyses of turbulent flows have been used in various contexts as a tool to probe, quantify, and explain the physics underlying phenomenological studies of observed structures, such as near-wall streaks [\citen{delAlamo2006linear}] and their role in the amplification of  streamwise vortices [\citen{schoppa2002coherent}], and hairpin structures [\citen{Sharma13}], and self-similar structures [\citen{Moarref13,sharma2017scaling}]. 

A particularly fruitful approach for operator-based modal analysis of turbulent wall-bounded flows, utilized in several of the aforementioned studies, comes from consideration of the resolvent operator associated with the mean-linearized equations [\citen{Farrell93,hwang2010linear,McKeon2010}], and in particular its singular value decomposition.  More generally, the utility of such analysis in fluids arises because the pseudospectrum is often more relevant than the spectrum for understanding typical instability and amplification mechanisms [\citen{ReddySchmidHenningson, reddy1993energy, Trefethen:Science93, Jovanovic05, symon2018normal}].  The nonlinear terms appear as a feedback interconnection with the linear resolvent operator within this analysis framework.  As such, a gain-based (input/output) decomposition of the linear resolvent operator provides insights regarding the amplification of velocity and pressure modes due to the nonlinear forcing terms~[\citen{McKeon2010}].  Recently, the resolvent formalism has been extended to study the influence of surface roughness effects, providing a convenient tool for studying passive flow control devices such as, e.g., compliant surfaces  [\citen{Luharcompliant15}] and spanwise periodic and streamwise-constant riblets~[\citen{chaverin2019}].  Similar extensions can also be used to design active control strategies [\citen{Luharcontrol14}].

Although the resolvent formalism has become a prominent method for analyzing wall-bounded turbulent flows in recent years, the approach has close connections with other modal analysis techniques as well.  Under the assumption that forcing results in uncorrelated resolvent response mode expansion coefficients, it can be shown that resolvent response coincide with spectral POD modes [\citen{towne2018spectral}], providing a connection between operator-based and data-driven modal decompositions.  Indeed, POD has a rich history in the study of  wall-bounded turbulent flows [\citen{Lumley1967,Lumley1981,Berkooz1993,Holmes96}], with the initial approaches being tractable using two-point correlation measurements [\citen{Lumley1967}].  

Recent investigations have also applied resolvent-based models for state estimation from limited measurements [\citen{Beneddinestep16,illingworth2018estimating}], and have also leveraged covariance completion techniques to model the nonlinear forcing terms within the resolvent framework as appropriate colored noise processes [\citen{zare2017colour,Towne2019}].  These recent investigations may offer a convenient set of reduced-complexity models that can guide future investigations on controlling turbulent wall-bounded flows.

\subsection{Spatially developing flows}
\label{sec:bl}

In contrast to fully developed parallel channel flow, the boundary layer over a flat plate grows slowly in the streamwise direction. The boundary layer thickness grows as $\delta \sim \sqrt{x\nu/U_\infty}$, where $U_\infty$ is the free stream velocity and $\nu$ is the kinematic viscosity.  At a sufficiently high Reynolds number, a disturbance generated at an upstream location grows in amplitude as it is transported downstream by the mean flow. Therefore, the flow is globally stable but is locally convectively unstable. The latter instability refers to the fact that a local stability analysis would yield an unstable system, based on a parallel flow assumption with the mean profile taken from a particular fixed streamwise position. In the full physical domain with a global viewpoint, however, the growth of perturbations at a fixed streamwise position is only a transient phenomenon, thus rendering the system as asymptotically stable [\citen{chomaz:05}]. 

The characteristic feature of convectively unstable flows is that they behave as amplifiers when externally forced. In particular, external perturbations (e.g., acoustic waves and free stream perturbations) continuously penetrate the boundary layer during a receptivity phase and trigger disturbances (Tollmien--Schlichting (TS) waves or streamwise vortices) that grow as they propagate downstream with the mean flow. If these disturbances reach above a certain threshold in amplitude, they may induce a breakdown to a turbulent flow.  The focus of a large number of studies has thus been on transition control that aim to delay the transition process by suppressing the growth of boundary-layer disturbances.   Within this context, modal decomposition techniques have been instrumental for reducing the number of degrees of freedom of the fluid system (typically $\gtrsim  \mathcal{O}(10^6)$) to yield a modal-based reduced-order model (ROM) (typically $\lesssim \mathcal{O}(10^2)$). Specifically, efficient and small reduced-order models can be constructed when the input-output dynamics are much simpler than the full spatiotemporal perturbation dynamics. For example, this is the case for feedforward control of TS waves using a few strategically-placed actuators and sensors flush-mounted on a flat plate. The output signal from an upstream sensor used to detect propagating disturbances is fed to a suitable controller, which in turn provides an actuation signal (input) that  attenuates the measured disturbances through interference. 

The construction of modal-based ROMs has been particularly successful using BPOD modes [\citen{bagheri:brandt:henning:09, 2009:barbagallo:sipp:schmid}]. A BPOD basis takes into account the  sensitivity to upstream forcing via the adjoint balanced modes. In contrast, the leading POD modes represent the most energetic structures located far downstream and thus have little spatial support upstream near the forcing. This makes it difficult to obtain small and accurate Galerkin models of the input-output dynamics. A number of studies have also used the direct and adjoint eigenmodes of the linearized system as an expansion basis to construct ROMs~[\citen{aakervik:hoepffner:ehrenstien:henning:07}]. However, these models quickly become ill-conditioned, as the streamwise separation between consecutive pairs rapidly increases~[\citen{bagheri:hoepffner:schmid:henning:09}]. Although modal-based ROMs have resulted in experimentally viable controllers~[\citen{Fabbiane:2015, Simon2016}], one of their limitations is that they require detailed knowledge of the spatial distribution of the upstream disturbance source (or noise environment). This requirement, which poses a limitation in experimental settings in particular, have resulted in a number alternative approaches to obtain ROMs based on system identification methods~[\citen{herve:2012}]. 

A number of groups have used modal decomposition techniques to extract and understand the inherent dynamics of spatially developing flows. However, extraction of temporal dynamics using data-driven methods is a challenging problem in noisy environments. If the system is continuously driven by external noise, the system will, after a transient, reach a statistically-steady state, thus causing the collected snapshot data to contain both external driving and inherent dynamics~[\citen{inigo:2016}]. A DMD analysis performed on such a data set will provide a spectrum with marginally stable eigenvalues, which is in contradiction with the damped spectrum of these systems. However, the DMD modes provide information about spatial inherent dynamics corresponding to a spatial stability analysis. 

\begin{figure}
\small
\centering
\begin{tabular}{ll}
    (a) & (b) \\
    \includegraphics[width=0.45\textwidth]{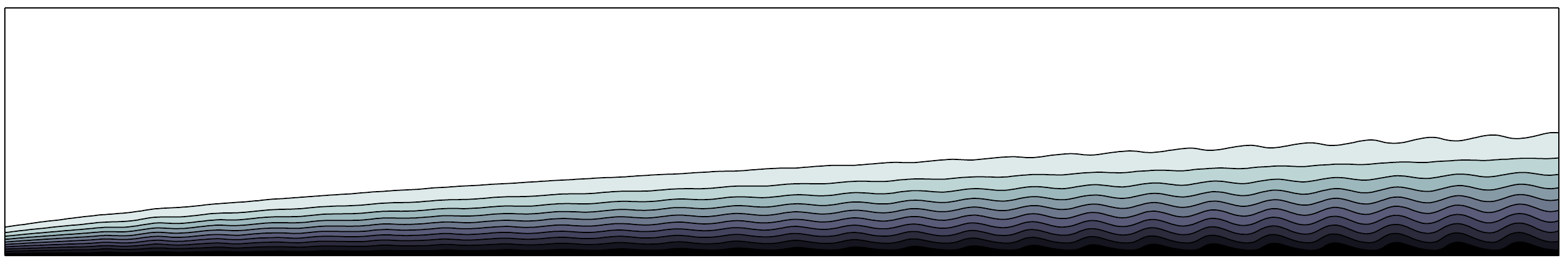}
    &
    \includegraphics[width=0.45\textwidth]{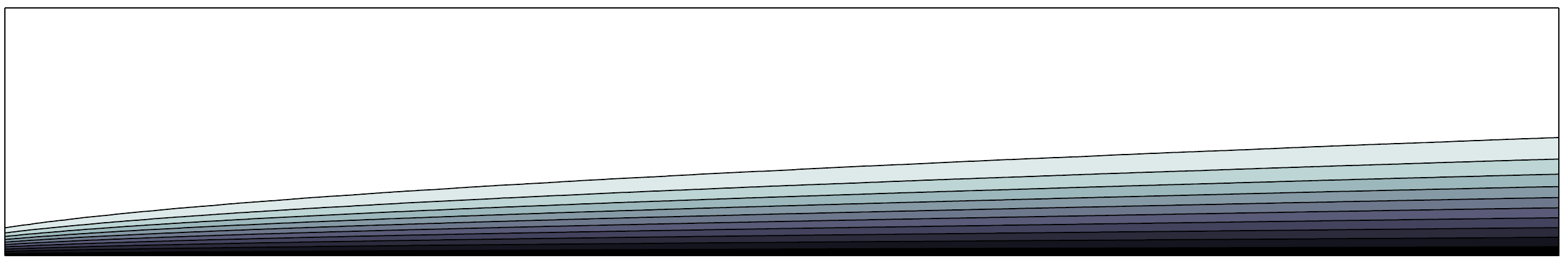}\\
    (c) & (d) \\
    \includegraphics[width=0.45\textwidth]{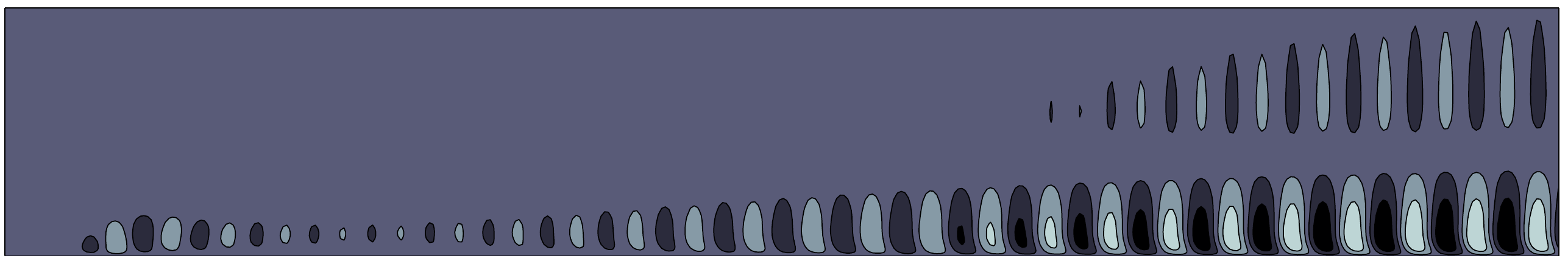} 
    &
    \includegraphics[width=0.45\textwidth]{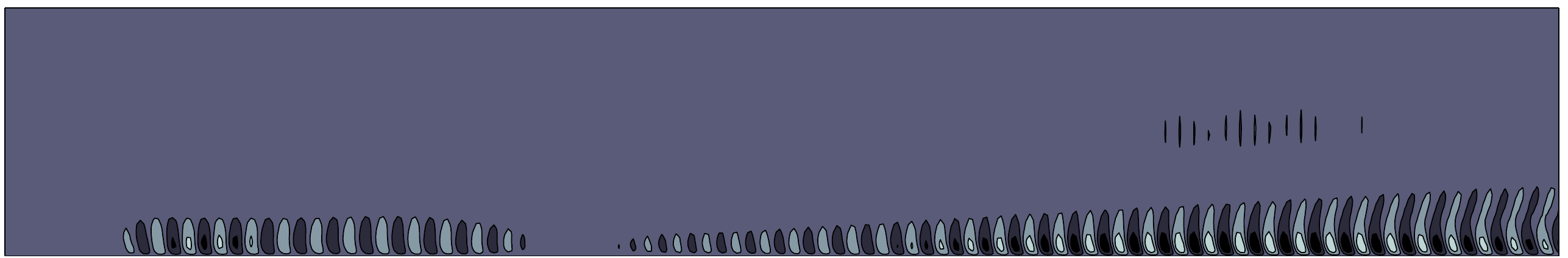}
    \end{tabular}
   \caption{(a) An instantaneous snapshot of the streamwise velocity component of the harmonically forced flat-plate boundary layer with non-dimensional frequency $F = 10^6 \omega \nu/U_\infty^2=120$. The inlet Reynolds number based on the displacement thickness and free-stream velocity is $Re_{x}=U_\infty x/\nu=30\,387$ in a two-dimensional computational box. The zeroth, first and third  DMD modes (all marginally stable) of the system are shown in frames (b-d) respectively. A TS wave is observed (c) that grows in the streamwise direction between branches I and II at a rate predicted by the local spatial stability analysis.}
   \label{fig:bl_DMD}
\end{figure}

To illustrate this, let us consider the uniform flow over a flat plate, where a localized harmonic forcing in the wall-normal direction of frequency $\omega$  is continuously applied upstream in the boundary layer.  
An instantaneous snapshot of the streamwise velocity component  is shown in Fig.~\ref{fig:bl_DMD}(a). The growing boundary layer is modulated by periodic forcing. 
The zeroth, first and third DMD modes are shown in Fig.~\ref{fig:bl_DMD}(b-d). The zeroth DMD mode corresponds to the time-averaged mean flow, which for this small-amplitude forcing is very close to the Blasius solution. The first DMD mode corresponds to a TS wave, where the associated DMD eigenvalue $\lambda_j$  has zero growth rate and a frequency that is equal to the forcing frequency. As mentioned earlier, the zeroth temporal growth rate predicted by DMD is in contradiction with a global spectrum of the system, which predicts damped system. However, in Fig.~\ref{fig:bl_DMD}(c), we observe that the amplitude of the DMD mode decays immediately downstream of the location of the forcing before it begins to grow at a particular streamwise location (branch I) until it peaks further downstream (branch II). The spatial locations of branches I and II for this particular forcing frequency corresponds to values obtained from the neutral curve of a local analysis of the Blasius boundary layer [\citen{Schmid01}].   The third DMD mode in Fig.~\ref{fig:bl_DMD}(d) corresponds to another TS wave with frequency $2\omega$ generated from nonlinear interactions.  

\section{Airfoil wakes}
\label{sec:airfoil}

Flow over an airfoil is another example that attracts great engineering interests in aerodynamic and turbomachinery applications.  Modal analysis has examined various aspects of airfoil wakes, including the wake structures [\citen{Wu:JFM1998, MGM:AIAAJ2017}], body geometry [\citen{LeGresley:AIAA2000, Thomareis:PoF2017}], tip vortex [\citen{Devenport:JFM96, DelPino:AJ2011, Edstrand:JFM2016, Edstrand:JFM2018b}], aeroacoustics [\citen{Tam:JFM2012, FosasSchmidSipp:JFM2014, Ribeiro:PoF2017}], and buffeting [\citen{Liepmann:JAS1955, Roos:AIAAJ1980, Raveh:AIAAJ2009, Crouch:JFM2009}].  Key efforts have been placed on mitigating flow separation over an airfoil for performance enhancement and improved safety of aircraft.  In this section, we discuss how modal analysis can be used to study the flow physics over the airfoil and how its insights can be utilized to develop effective separation control strategies.  
In Sections IV annd V, we focus on the {\it excitation} of modes to modify the mean flow profile.  Although, we can also consider the {\it suppression} of modes as a way to control fluid flows, its effectiveness on modifying the flow can be in question for higher Reynolds number flows, in which nonlinear effects are strong.  On the other hand, excitation can push the flow away from its current state, if successful, and alter the mean flow profile across a range of Reynolds numbers as some of the examples that follow will show.

\subsection{POD and DMD analyses}

We have discussed the importance of the time window and temporal resolution for the snapshots used in the data-based modal analyses.  For high-Reynolds-number separated flows, high temporal resolution is required to capture the shear-layer structures over the airfoil.  To accurately capture the wake structure, on the other hand, we need to ensure that the snapshots are adequately collected over a reasonable number of vortex-shedding periods.  For turbulent flows, a large number of snapshots in time are needed due to the chaotic nature of the wake dynamics.

The spatial domain for data-based analyses should be based on the physics under examination.  If global snapshots of a high-Reynolds-number separated flow are used to perform POD, shear-layer structures over the separation bubble may only be revealed at high-order modes.  This is because POD modes are ranked with respect to the relative energy content and shear-layer structure usually contain a smaller fraction of energy compared to the wake structure.  If the shear-layer structure is of the main interest, we can consider the domain to cover only the separation bubble so that the shear-layer structures can be analyzed with low-rank POD modes.  We also note that the same purpose can be served by introducing the spatial window as a weighted function.  If DMD is performed, it can automatically separate the shear-layer structure from the wake structure according to their own corresponding frequencies, as each DMD mode holds a single frequency. 

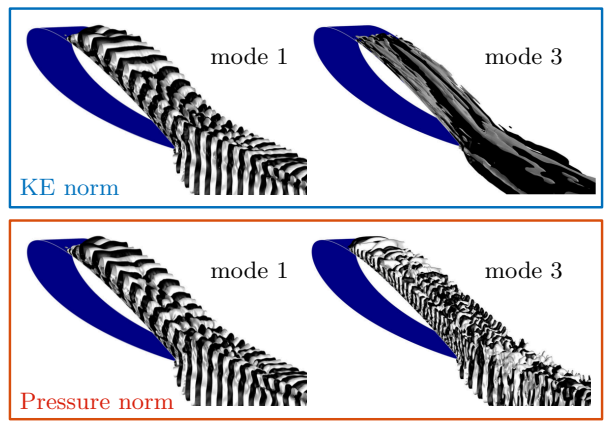
\begin{figure}
\centering
    \begin{overpic}[width=0.48\textwidth]{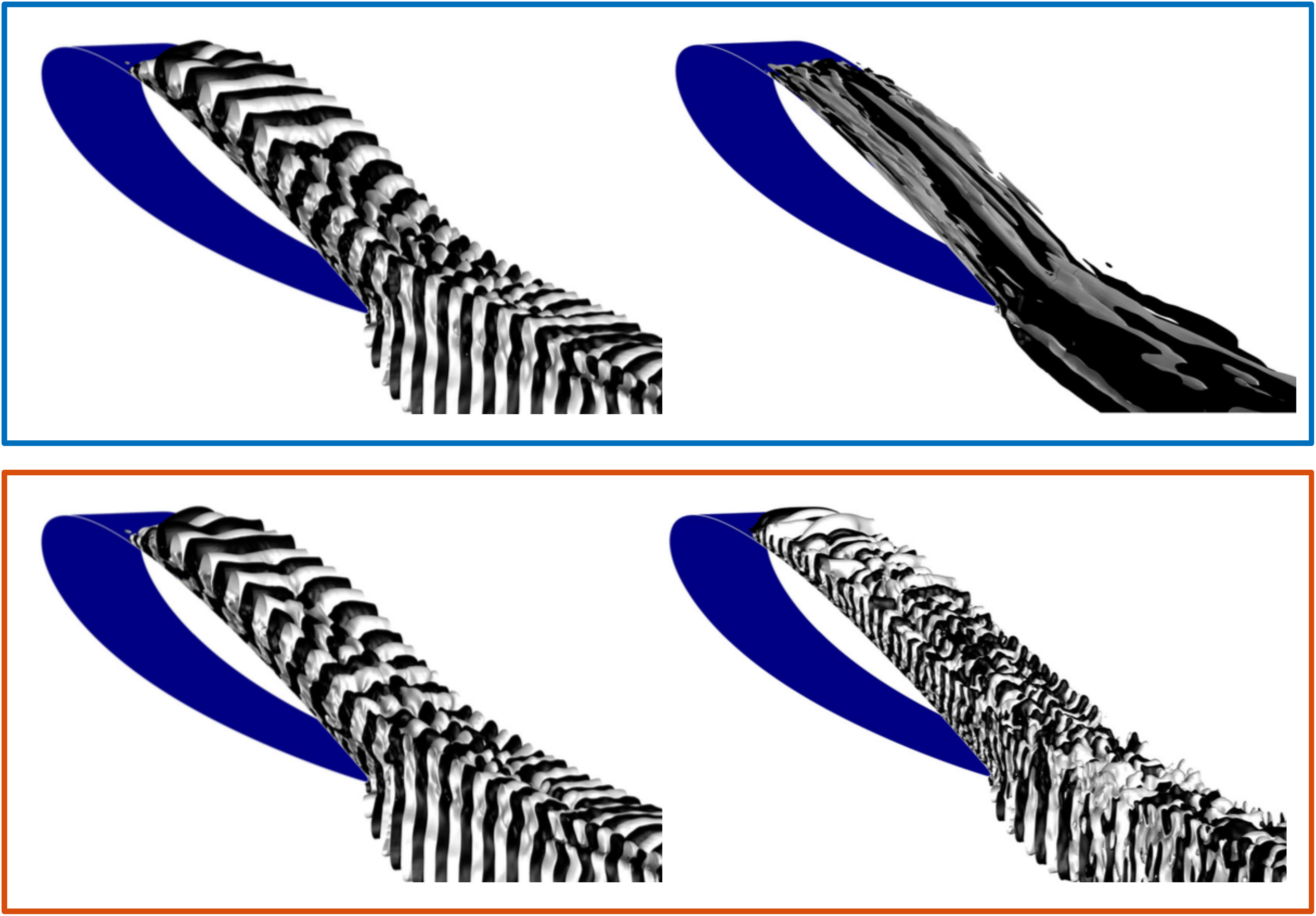}
    	\small
    	\put(2,38){\color{blue}KE norm}
		\put(2,2){\color{red}Pressure norm}
		\put(34, 60){mode 1}
		\put(80, 60){mode 3}
		\put(34, 24){mode 1}
		\put(80, 24){mode 3}
	\end{overpic}
	\caption{POD modes obtained with perturbation kinetic-energy (KE) norm (top) and pressure norm (bottom) [\citen{Ribeiro:PoF2017}] reveal distinctive structures in the second mode pair (mode 3). Reprinted with permission from AIP Publishing.}
	\label{fig:RibeiroWolf_2017}
\end{figure}

Data-based modal analysis on airfoil flows has been shown to be capable of capturing coherent structures at chord-based $Re$ up to $\mathcal{O}(10^5)$ [\citen{Mohan:JoA2017,Thomareis:PoF2017}].  The studies by Wolf et al.~[\citen{Ribeiro:PoF2017, Ricciardi:AIAA2019naca}] consider a 3D dataset collected for the turbulent flow over a NACA 0012 airfoil at $Re = 408\,000$.  Similar to Freund and Colonius [\citen{Freund:IJA2009}], their POD analysis considers the use of different norms to reveal the flow structures that are associated with tonal noise.  The leading POD modes using the norms based on kinetic energy and pressure fluctuation are shown in \fig \ref{fig:RibeiroWolf_2017}.  The use of these norms reveals similar structures associated with the generation of dominant tonal noise according to the spectral content of their temporal coefficients.  However, the POD modes from the second pair (mode 3) exhibit different structures with the use of different norms.  While the use of the pressure norm uncovers spanwise structures associated with the harmonics of the dominant tone, the kinetic-energy norm reveals streamwise structures over the airfoil which do not attribute to the tonal noise generation.  This study highlights the importance of the choice of norms in POD analysis.  It also suggests that the collection of 3D dataset can be necessary even when spatial homogeneity may appear appropriate, as the energetic streamwise structure in mode 3 would not have been revealed if only a spanwise slice of data was considered in the modal analysis. 

\subsection{Global stability analysis}

\begin{figure}
\centering
    \begin{overpic}[width=0.97\textwidth]{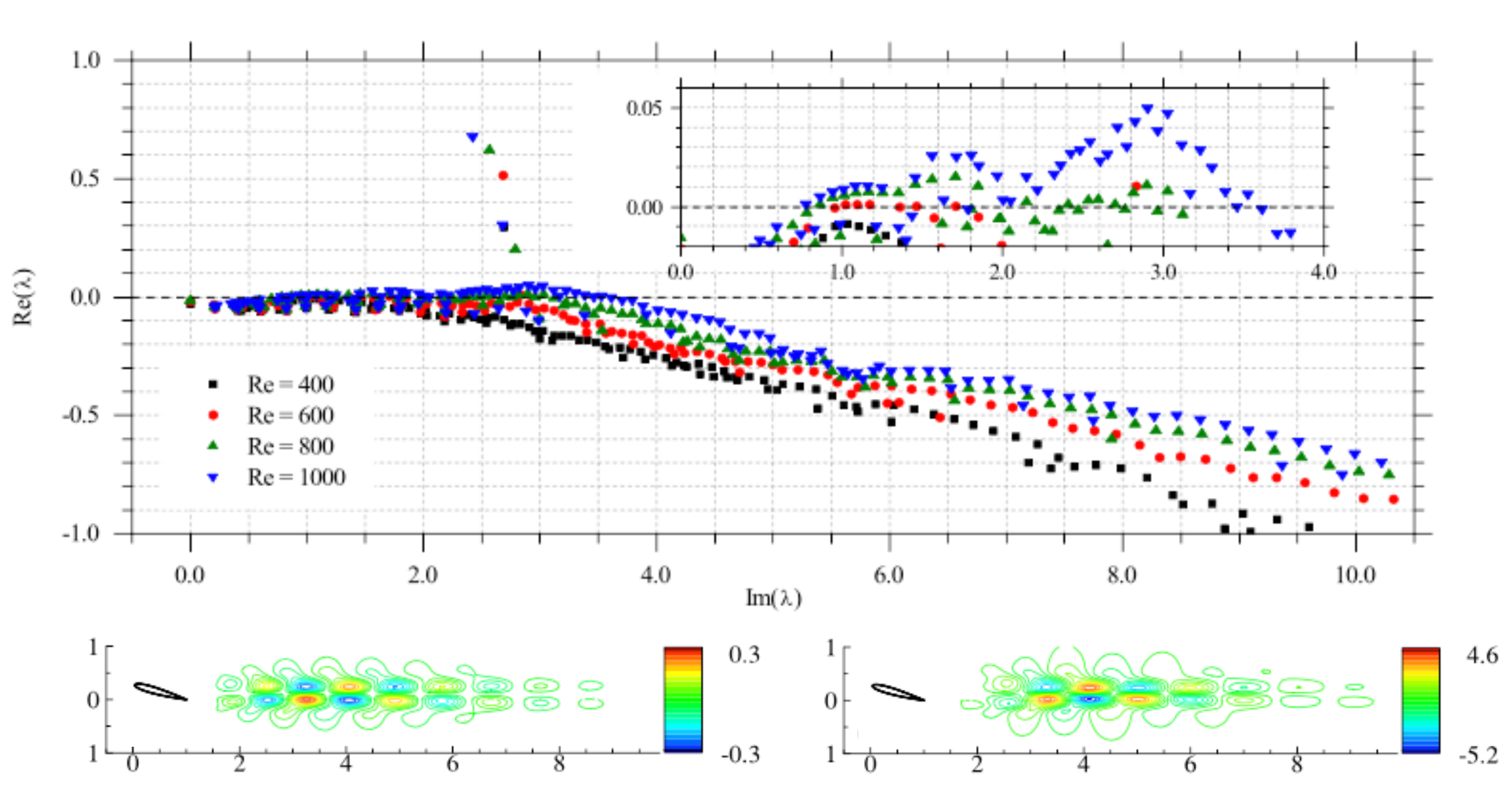}
    \small
		\put(9, 9){$Re = 600$}
		\put(58, 9){$Re = 1000$}
	\end{overpic}
	\caption{The spectrum (left) and dominant unstable eigenmodes (right) from the biglobal stability analysis on flows over a NACA 0012 airfoil [\citen{Zhang:PoF2016}].  Results from the spanwise wavenumber $\beta = 0.0001$ are shown.  The streamwise velocity components of the eigenmodes are shown for $Re = 600$ and $1\,000$. Reprinted with permission from AIP Publishing.}
	\label{fig:ZhangSamtaney_2016}
\end{figure}

Global stability analysis of flows over a NACA 0012 airfoil has been performed by Theofilis [\citen{Theofilis:AIAA02}] and Zhang and Samtaney [\citen{Zhang:PoF2016}].  An example of the spectrum and eigenmodes obtained from Zhang and Samtaney [\citen{Zhang:PoF2016}] is presented in \fig \ref{fig:ZhangSamtaney_2016}.  Here, unstable equilibrium flows over the airfoil at $Re = 400$ to $1\,000$ are considered as the base states.  With the spatial periodicity assumed in the spanwise direction, they adopted the biglobal mode representation $\boldsymbol{q}'(\boldsymbol{x}) = \hat{\boldsymbol{q}}(x, y) \exp(i \omega t + i \beta z)$, where $z$ is the homogeneous spanwise direction and $\beta$ is the spanwise wavenumber.  Unstable eigenvalues are found for the base flows at all selected Reynolds numbers, corresponding to the unstable vortex shedding observed in the companion direct numerical simulations.  Moreover, the dominant unstable eigenmodes reveal vortex-shedding structures in the pattern of bluff body wake.  These corresponding frequencies determined from the stability analysis agree with the shedding frequencies observed in the direct numerical simulations.  The destabilizing effect from increasing Reynolds number is also reflected in the increasing growth rates of the eigenvalues. 

\begin{figure}
\centering
    \begin{overpic}[width=0.48\textwidth]{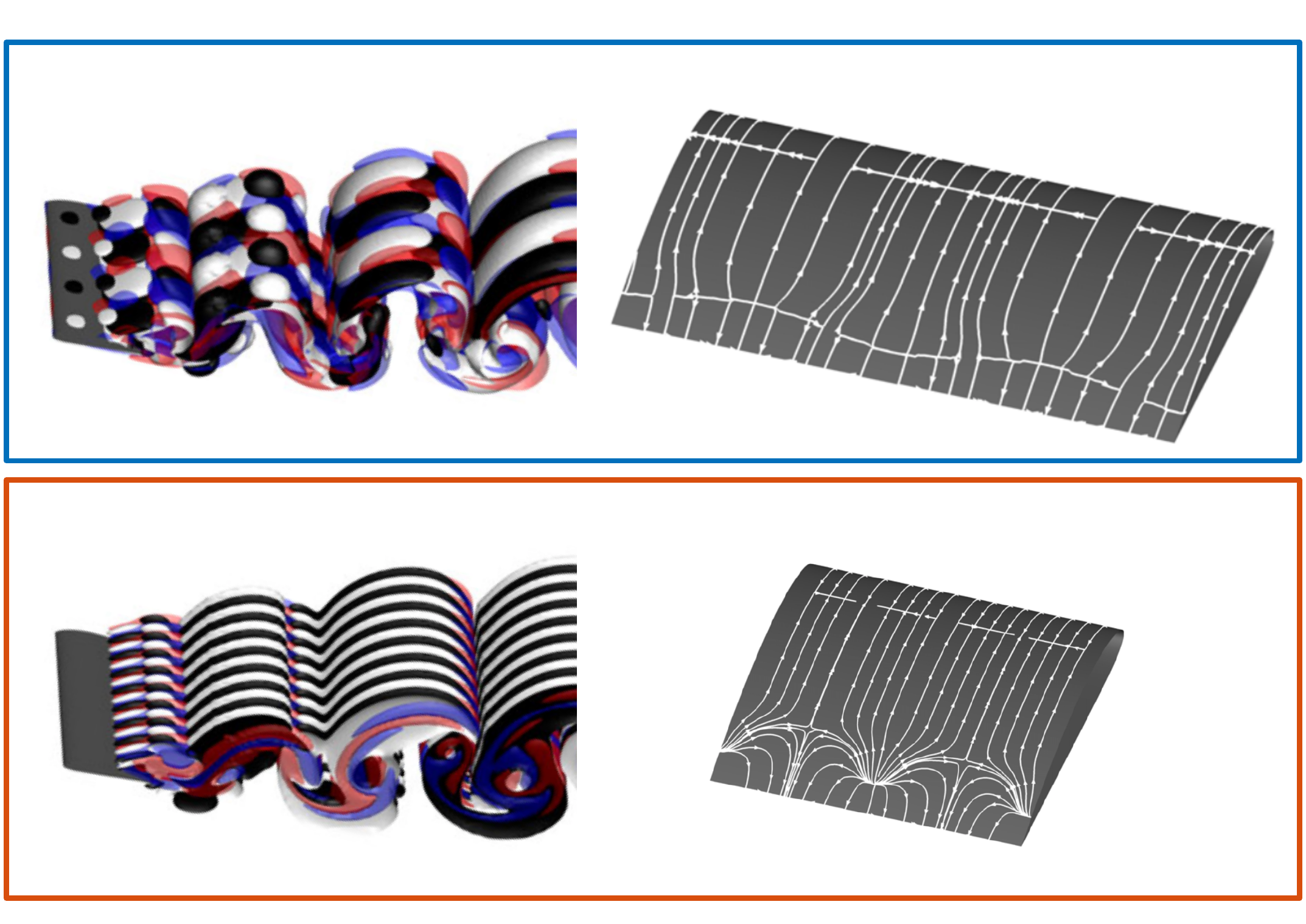}
    	\small
        \put(2,63){\color{blue}long-wavelength mode ($\beta = 3$)}
        \put(2,29.5){\color{red}short-wavelength mode ($\beta = 11$)}
    \end{overpic}
	\caption{Floquet analysis of the periodic flow over a NACA 4415 airfoil at $Re = 500$ and $\alpha = 20^\circ$ reveals two 3D instability modes at $\beta = 3$ (top) and $11$ (bottom) [\citen{He:JFM2017}]. Modal structures and surface streamlines are shown for each mode. Reprinted with permission from Cambridge University Press.}
	\label{fig:HeEtal_2017}
\end{figure}

Global stability analysis can be extended to periodic base flows by incorporating Floquet analysis [\citen{Barkley:JFM1996}].  The stability of the periodic base flows representing the vortex shedding in airfoil wake has been examined with Floquet analysis by He \etal~[\citen{He:JFM2017}].  In their study, the 3D instability is treated as a secondary instability that takes place about the 2D periodic base flow.  The stability is determined by the magnitude of the Floquet multiplier, which indicates the growth/decay rate of the 3D perturbation when propagating with the 2D periodic base flow.  Considering the periodic base flow over the NACA 4415 airfoil at $Re = 500$, two instability modes, as shown in \fig \ref{fig:HeEtal_2017}, appear at the spanwise wavenumbers $\beta = 3$ and $11$, exhibiting distinct surface flow patterns.  The short-wavelength instability ($\beta = 11$) is the stronger of the two, producing the wall-shear distributions pattern that portrays a 3D flow pattern. 

\subsection{Flow control}

\subsubsection{Separation control using resolvent analysis}

An airfoil at a high angle of attack or in unsteady maneuver can experience stall resulting from flow separation over the suction surface [\citen{Wu:JFM1998, Carr:JA88, Garmann:PF11, Jantzen:PF14, Benton:JFM19}].  To address this issue, development of separation control technique to suppress flow separation has been the focus of many studies [\citen{Wu:JFM1998, Greenblatt:PAS00, Amitay:AIAAJ2002, Munday:AIAAJ2018}].  The study conducted by Yeh and Taira [\citen{Yeh:JFM2019}] considered the use of resolvent analysis to guide active separation control with periodic forcing.  The analysis was conducted about the turbulent mean flows obtained from the baseline (uncontrolled) simulation to determine the optimal actuation frequencies and wavenumbers to suppress separation.  As a precursor to resolvent analysis, global stability analysis was conducted, which revealed asymptotic instability of their base flows.  To extend the resolvent analysis to the unstable base flows, the finite-time (discounted) analysis [\citen{Jovanovic:Thesis2004}] was applied by selecting a complex frequency $\omega = \omega_r - i\alpha$ in the resolvent operator $H(\omega) = \left[i\omega I - L_{\overline{\boldsymbol{q}}}\right]^{-1}$.  The real-valued discounting parameter $\alpha$ was chosen to be higher than the dominant unstable modal growth rate of $L_{\overline{\boldsymbol{q}}}$ from the companion global stability analysis such that the energy amplification in the input-output analysis is examined over a shorter time scale than that of the dominant instability.  

\begin{figure}
\centering
	\includegraphics[width=0.95\textwidth]{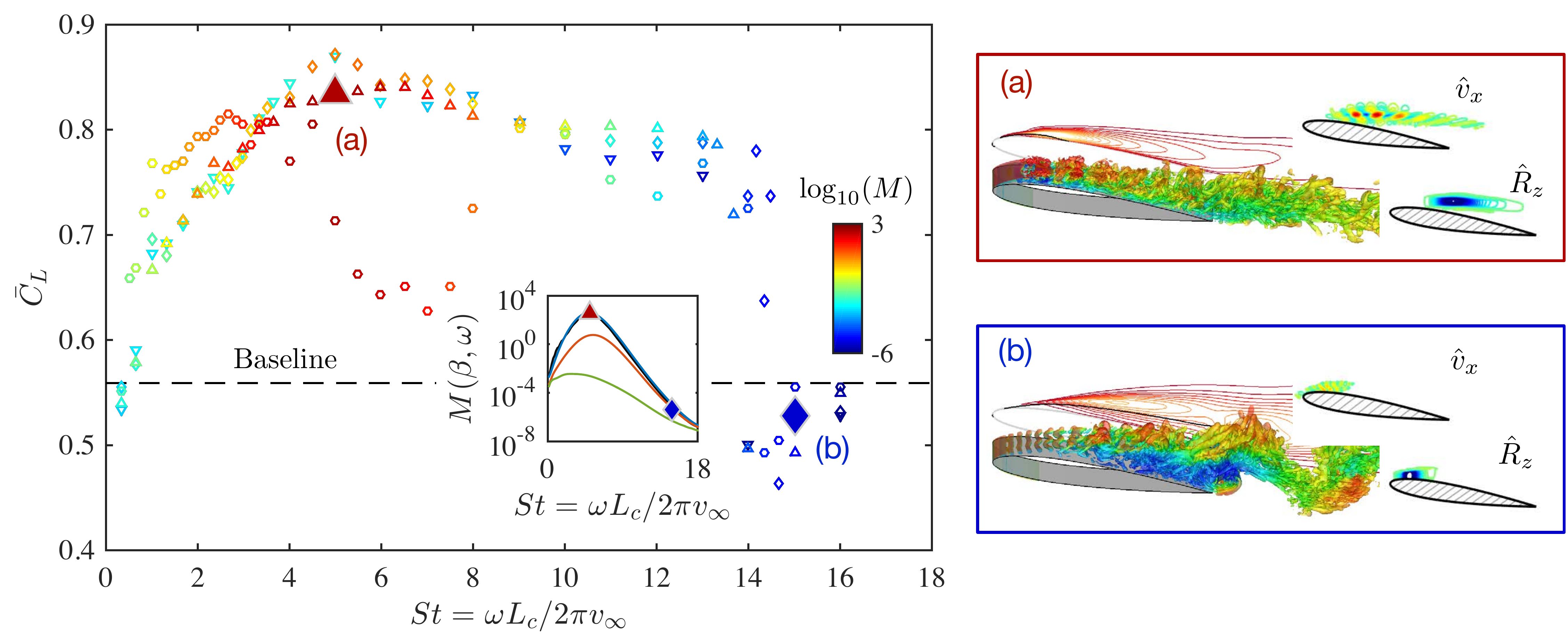}
	\caption{The use of resolvent analysis to develop separation control for an airfoil wake at $Re = 23\,000$ [\citen{Yeh:JFM2019}]. Control effectiveness is well predicted by the modal mixing function $M(\beta, \omega).$  For cases (a) and (b), visualizations are shown for the instantaneous flow fields and the resolvent response modes (streamwise velocity component) and the spanwise modal Reynolds stress $\hat{R}_z = Re(\hat{v}_x^*\hat{v}_y)$.}
	\label{fig:YehTaira_2018}
\end{figure}

Flow separation can be suppressed by the entrainment of free-stream momentum over the suction surface, which can be achieved by enhancing the momentum mixing.  As such, Yeh and Taira [\citen{Yeh:JFM2019}] proposed a modal mixing metric $M(\beta, \omega) \equiv \int_\Omega [ \sigma^2 (\hat{R}_x^2 + \hat{R}_y^2 + \hat{R}_z^2)^{\frac{1}{2}}]_{\beta, \omega} w(\boldsymbol{x}) {\rm d}\boldsymbol{x}$ that incorporates the response modal Reynolds stresses $\hat{R}$ and the associated gain $\sigma$ from resolvent analysis, with the spatial weight function $w(\boldsymbol{x})$ that encompasses the shear layer region over the suction surface.  This scalar metric $M$ is a function of the spanwise wavenumber $\beta$ and frequency $\omega$ and quantifies the momentum mixing that takes place over the separation bubble.  With enhanced free-stream entrainment, the scalar function $M(\beta, \omega)$ assesses the effectiveness of the choice $(\beta, \omega)$ in suppressing flow separation.  Supported by the independent parametric study on open-loop controlled flows,  the modal mixing metric $M(\beta, \omega)$ was found to predict the enhancement of aerodynamic performance over the actuation parameter space of $\beta$ and $\omega$.   The agreement between $M(\beta, \omega)$ and the performance enhancement is presented in \fig \ref{fig:YehTaira_2018}, where two cases are highlighted by showing the resolvent modes and visualization of instantaneous flow field.  In case (a), high amplification and strong modal Reynolds stresses over the suction surface give rise to the peak magnitude of the modal mixing metric $M(\beta, \omega)$.  Companion LES confirms high level of momentum mixing induced by flow control using actuation with these parameters of $(\beta, \omega)$, which attaches the flow and achieved enhancement in both lift and drag.  In contrast, poor performance enhancement for case (b) with insufficient mixing provided by the small roll-up structure is also suggested by the resolvent response mode and low level of $M(\beta, \omega)$.  This example of using modal analysis provides quantitative assessments of the control effectiveness over parameter space without the need for a computationally expensive LES parameter study.

\subsubsection{Attenuation of wing-tip vortex}

Another important feature of an airfoil wake is the tip vortex.  A finite-span wing at an angle of attack creates a tip vortex due to the pressure difference between the suction and pressure surfaces.  Since tip vortices have implications for efficiency loss, safety concerns, and wake unsteadiness, there have been a number of efforts to characterize the tip vortex [\citen{Devenport:JFM96, Crouch:CRP05, Bailey:JFM08}].  Flow unsteadiness and instabilities around a wing tip have been studied with modal analysis as well.  In the experimental work of Edstrand et al.~[\citen{Edstrand:JFM2016}], POD analysis of the wake on cross-stream planes (from PIV measurements at $Re = 530\,000$ and $\alpha = 5^\circ$) revealed modes associated with the so-called wandering phenomena of the tip vortex.  The modal structures from POD were compared to the stability modes obtained for a Batchelor vortex model, exhibiting great similarity.  This observation suggests the wandering phenomenon to be closely related to the tip vortex instability.

\begin{figure}
\centering
	\includegraphics[width=0.99\textwidth]{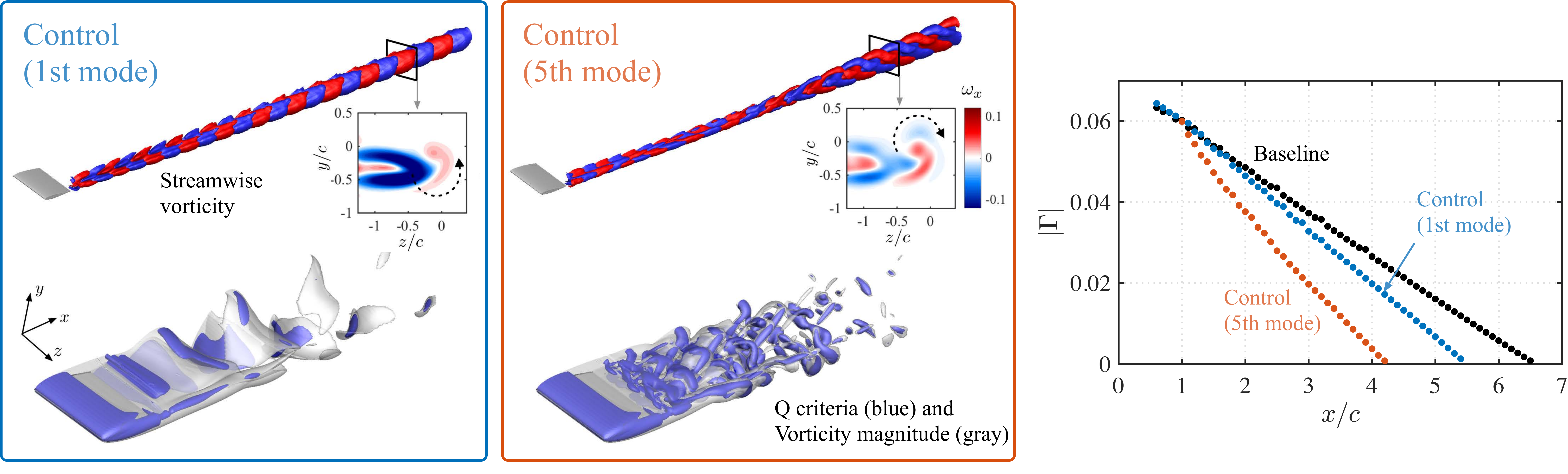}
	\caption{The first and fifth instability modes visualized (top).  The modal profiles are used to introduce perturbations near the trailing edge to modify the wake and attenuate the tip vortex (bottom).  Circulation of the tip vortex is compared (right) to assess the effectiveness of active flow control. Compiled with permission from Edstrand et al.~[\citen{Edstrand:JFM2018b}].  
	Reprinted with permission from Cambridge University Press.
	}
	\label{fig:EdstrandEtal_2018b}
\end{figure}

The prolonged presence of tip vortices is a safety hazard for aircraft operations.  To address this issue, there have been efforts to attenuate the tip vortices with active flow control.  The traditional approach introduces perturbations from the wing tip in hopes of weakening the tip vortex  [\citen{Margaris:2010,Edstrand:2015}].  In a recent study by Edstrand et al.~[\citen{Edstrand:JFM2018b}], the wake behind the wing tip was computationally analyzed via the global stability analysis with a parabolized formulation incorporated in the streamwise direction ($Re = 1\,000$ and angle of attack of $5^\circ$).  Similar approach was taken in an earlier study of the 3D stability characteristics of an elliptic wing wake by He et al.~[\citen{He:TCFD17}].  In the detailed analysis of Edstrand et al.~[\citen{Edstrand:JFM2018b}], they revealed two distinct types of instabilities, as visualized in \fig \ref{fig:EdstrandEtal_2018b}.  The dominant modes were found to possess structures that co-rotate with the tip vortex.  They also found the subdominant fifth instability mode that emanates from the trailing edge with structures that counter-rotate with the tip vortex.  Since past studies have revealed that counter-rotating instability modes can effectively attenuate vortices in freespace, this finding suggested the forcing input to be introduced from the trailing edge to attenuate the tip vortex, instead of the wing tip.  Using direct numerical simulation, the tip vortex was found to be weakened effectively with the fifth mode based control setup, as shown by the circulation of the tip vortex in \fig \ref{fig:EdstrandEtal_2018b}.  To ensure that the control technique does indeed trigger the counter-rotating instability, DMD was also used to assess the controlled flow, for which the expected counter-rotating perturbations was observed.  This study shows that detailed stability analysis can be used to design an effective control technique and identify the appropriate placement of the forcing input.

\section{Cavity flows}

\begin{figure}[hbpt]
    \centering
    \includegraphics[width=0.48\textwidth]{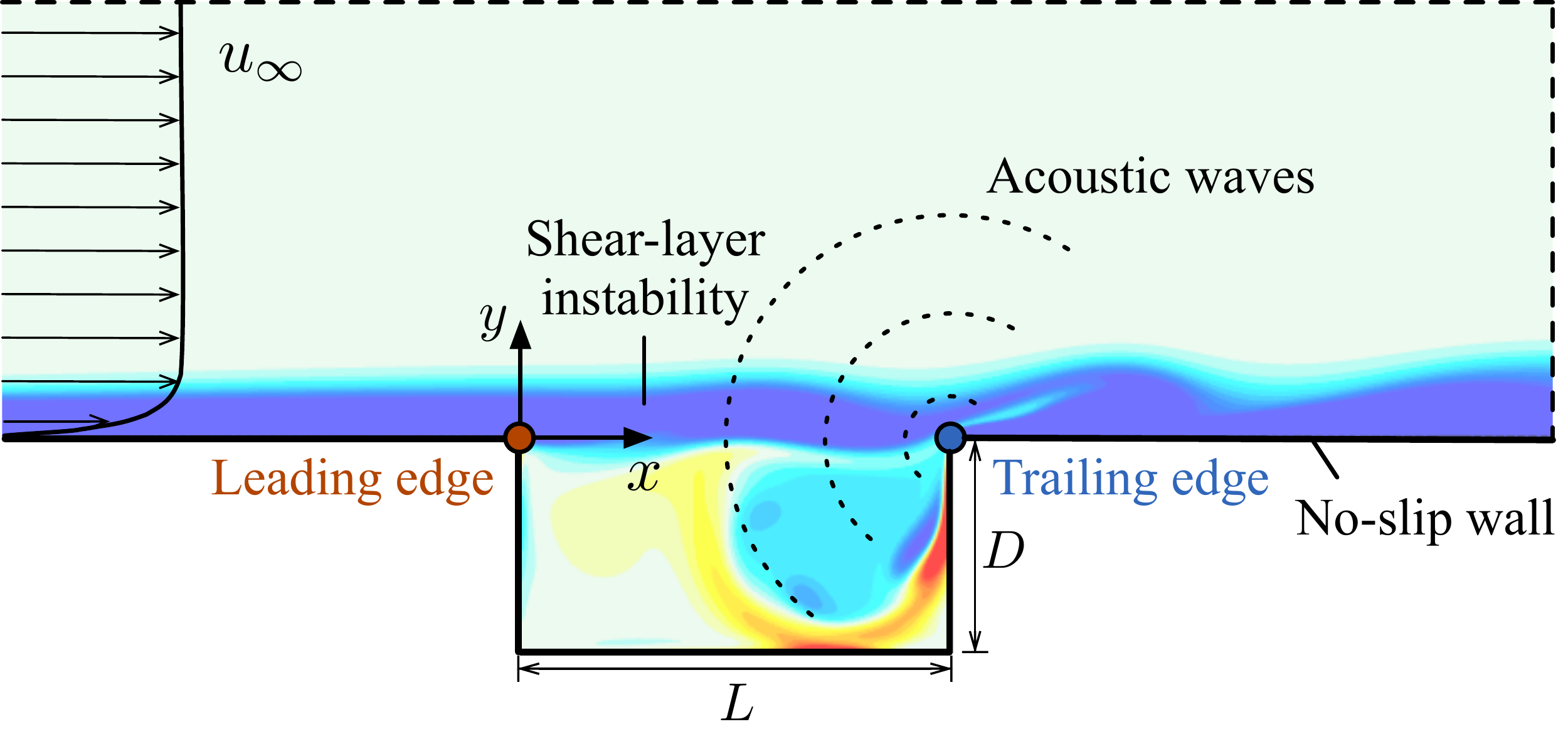}
    \caption{Schematic of open-cavity flow with spanwise vorticity in the background.}
    \label{fig:cavity}
\end{figure}

Flows over rectangular cavities serve as fundamental models for flows over landing-gear wells and stores on aircraft [\citen{Lawson:PAS11}].    
In an open-cavity flow, a shear layer forms from the leading edge and amplifies disturbances through the Kelvin--Helmholtz instability, which leads to the formation of large vortical structures that impinge on the cavity aftwall. Large pressure fluctuations and acoustic waves are generated from the impingement, which perturb the upstream shear layer.  This overall process forms a self-sustained natural feedback loop in open-cavity flows, as shown in \fig \ref{fig:cavity}. Resonant tones are generated from this process, which are known as the {\it Rossiter modes} [\citen{Rossiter:ARCRM64}]. 
Based on a large collection of experimental data, Rossiter derived a semi-empirical formula to predict the resonant frequencies.  As the original formula only considers the influence of the two important parameters of the freestream Mach number $M_\infty$ and cavity aspect ratio $L/D$, extensive experimental and numerical studies have followed to examine the influence of other parameters, including the Reynolds number and boundary layer thickness [\citen{Ahuja:NASA95,murray2009,Beresh:JFM16}].  

The spatial structures that correspond to the dominant Rossiter frequencies as well as coherent structures that arise from other types of instabilities or physics can be revealed by modal analysis techniques.
This section considers the applications of modal analysis techniques on rectangular open-cavity flows.  We further discuss ways to design flow control techniques to suppress flow oscillations based on insights from these modal analyses.  In what follows, we will focus on spanwise-periodic rectangular cavity flows, unless otherwise noted.

\subsection{POD and DMD analyses}
\label{cav_dmd}

Given a collection of snapshots of open-cavity flows from experiments or simulations, we can perform data-based analysis with POD and DMD.  As discussed in Section \ref{sec:cylinder} on cylinder flow, performing POD and DMD analyses requires a proper collection of snapshots to capture specific modes. For laminar cavity flow, the appropriate number of snapshots and length of time-series data can be estimated based on the fundamental Rossiter frequencies given by the semi-empirical formula. For a turbulent cavity flow, a large number of snapshots become necessary as such flow possesses spectra with a broader frequency content.   

POD analysis captures modes with large unsteady fluctuations.  
An example of applying POD analysis of turbulent flow over a cavity of $L/D = 6$ for $0.19 \le M_\infty \le 0.73$ from experiments has been briefly presented in our previous overview paper [\citen{Taira_etal:AIAAJ17}] (see Section III.B.2 and Murray \etal~[\citen{murray2009}]). Snapshots were collected using particle image velocimetry from high-speed flows. For compressible cavity flows, the spatial structures of the most energetic POD modes reside in the shear-layer region, showing the spatially growing nature of the modes towards the cavity trailing edge [\citen{murray2009}], which remain similar regardless of the free stream Mach number.  The POD analysis offers a framework to extract dominant energetic structures and serve as a foundation for systematic comparison over a range of operating conditions.

\begin{figure}[hbpt]
   \centering
   \includegraphics[width=0.99\textwidth]{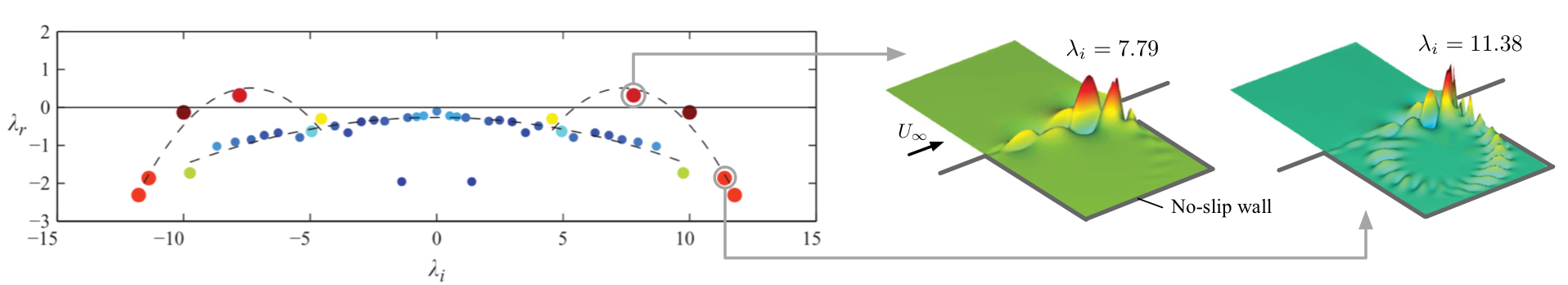}
   \caption{DMD eigenspectrum ($\lambda_i$: frequency, $\lambda_r$: growth/decay rate) for incompressible flow over cavity of $L/D=1$ and $Re=4\,500$ (figure adapted from [\citen{Schmid:JFM10}]). Inserted are streamwise velocity of dynamic modes. The size and color of eigenvalues represent the coherence measurement of each mode with respect to its structure size and energetic level. Reprinted with permission from Cambridge University Press.}
   \label{fig:DMD_Schmid}
\end{figure}

We can alternatively use the DMD analysis to extract dynamically important modes. If the flow field from a linearized Navier--Stokes solver is considered, the DMD analysis can return the global stability modes, as performed for linearized flow over a cavity of $L/D=1$ at $Re=4\,500$ by Schmid [\citen{Schmid:JFM10}].  Such analysis identifies the Kelvin--Helmholtz instabilities in the shear layer that cause flow oscillations.  As shown in \fig \ref{fig:DMD_Schmid}, the branch of DMD eigenspectrum containing unstable modes ($\lambda_r>0$) corresponds to shear-layer instabilities, as their modal structures are concentrated in the shear-layer region spanning the length of the cavity.  Instead of performing DMD on linear snapshots, regressions can also be applied to extract global instability modes. Br\`es and Colonius [\citen{Bres:JFM08}] conducted extensive linearized simulations and extracted 3D instabilities of compressible open-cavity flow using a regression approach.  

\subsection{Global stability and resolvent analyses} 
\label{cav_GSA}

Because open-cavity flow is globally affected by the Rossiter modes, biglobal stability analysis is widely used to study the perturbation dynamics about a given 2D base flow.  Here, let us consider the perturbations to be spanwise-periodic with a spanwise wavenumber of $\beta$ (normalized by cavity depth). The biglobal stability analysis assumes a homogeneous direction which we take to be periodic for the present discussion. If sidewall effects or any other 3D factors need to be considered, triglobal stability analysis can be adopted.  For the biglobal stability analysis with $\beta=0$, 2D eigenmodes can be found, which are closely related to the well-known Rossiter modes, and 3D instabilities can be determined by choosing $\beta>0$ [\citen{Yamouni:JFM13,Vicente:JFM14,Sun:TCFD16,Sun:JFM17}].  

\begin{figure}
   \centering
   \includegraphics[width=0.8\textwidth]{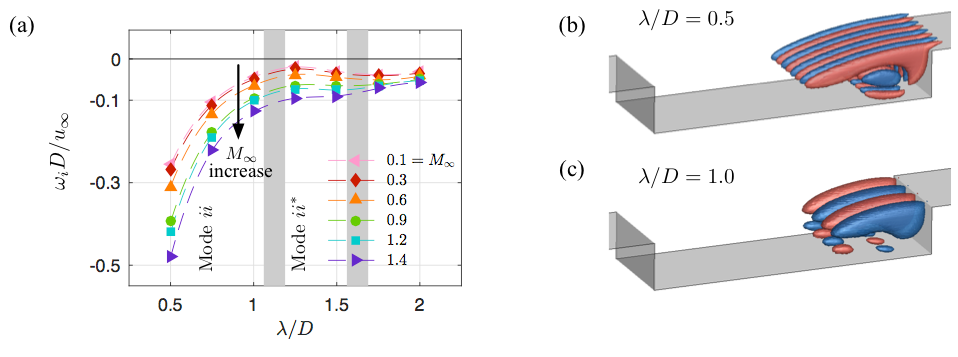}
   \caption{Biglobal stability results of compressible flows over cavity with $L/D=6$ at $Re_D=502$ and $M_\infty\in[0.1,1.4]$ (figure adapted from [\citen{Sun:JFM17}]). (a) Growth rate $\omega_i D/u_\infty$ versus spanwise wavelength $\lambda/D=2\pi/\beta$. (b) and (c) spanwise velocity iso-surfaces of the leading eigenmodes at $M_\infty=0.9$. Reprinted with permission from Cambridge University Press.}
   \label{fig:BiG_Sun}
\end{figure}

The insights gained from modal analysis can be leveraged to develop active flow control techniques to reduce the high-amplitude fluctuations in cavity flows.  Sun et al.~[\citen{Sun:JFM17}] performed biglobal stability analysis of laminar compressible flows over a long cavity of $L/D=6$ and $Re_D=502$ to derive physics-based  techniques for the attenuation of unsteady oscillations.  In their study, they examined the influence of Mach number $M_\infty$ and spanwise wavelength $\lambda/D$ on the 3D instability properties.  
It was found that the frequencies associated with 3D instabilities ($\beta > 0$) are one order of magnitude lower than those of the 2D shear-layer instabilities. The characteristics of the leading eigenmodes over a range of $M_\infty$ and $\lambda/D$ can be found as shown in \fig \ref{fig:BiG_Sun} (a). Increase in Mach number stabilizes the 3D leading eigenmodes. Furthermore, they noticed that the overall trend of growth rate $\omega_i D/u_\infty$ remain similar over the range of Mach numbers considered. This suggests that flow control strategies we design based on 3D instabilities may work across such range of Mach numbers.  Shown in Fig.~\ref{fig:BiG_Sun} (b) and (c) are the iso-surfaces of the leading eigenmodes with $\lambda/D=0.5$ and $1$. The 3D eigenmodes stem from centrifugal instabilities and are mainly concentrated in the rear part of the cavity where the recirculation zone resides. We can envision triggering the 3D instabilities to potentially remove kinetic energy from the shear layer and transfer them into the cavity.  Taking advantage of the three-dimensional flow instabilities is an efficient way to modify the flow field, as we shall see below in section V.\ref{cav_control}.  

\begin{figure}
   \centering
   \includegraphics[width=1.0\textwidth]{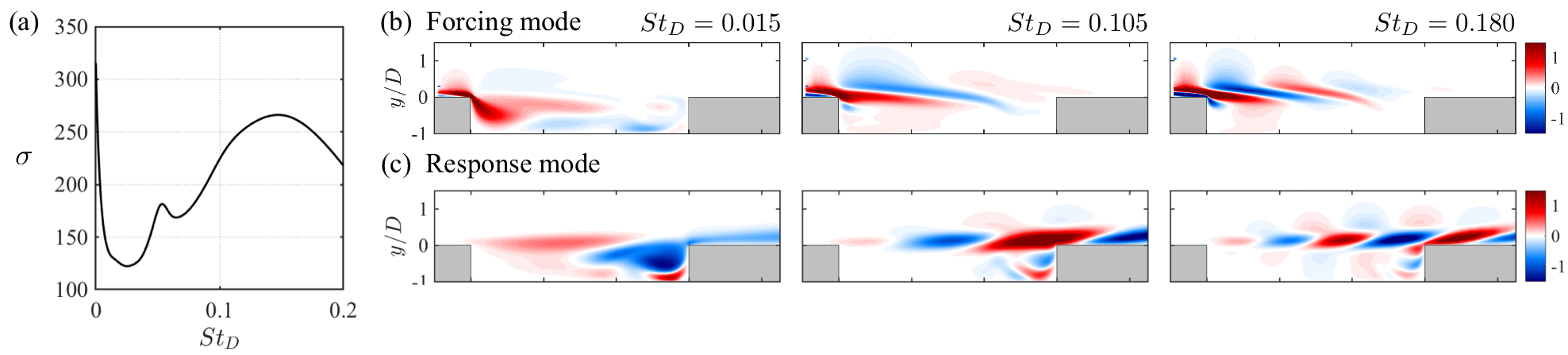}
   \caption{Resolvent analysis of compressible cavity flow of $L/D=6$, $Re_D=502$, $M_\infty=0.6$, and $\beta=2$. (a) Optimal amplification gain, (b) optimal forcing modes, and (c) optimal response modes. Contours of the streamwise velocity are visualized.}
   \label{fig:resl_mode}
\end{figure}

Resolvent analysis is capable of examining flow frequency response to harmonic forcing with respect to its base state. The optimal forcing and response modes can be identified according to their amplification gain $\sigma$ obtained by performing singular value decomposition of a resolvent operator $[i\omega I - L_{\bar{\boldsymbol q}}]^{-1}$. Shown in Fig.~\ref{fig:resl_mode} are some results of resolvent analysis of laminar flow over a cavity of $L/D=6$ at $Re=502$ and $M_\infty=0.6$ by Liu et al.~[\citen{Liu:AIAA18}]. By sweeping over a normalized frequency 
$St_D= \omega D/(2\pi u_\infty)$, a maximum gain appears around $St_D\approx 0.15$ (Fig.~\ref{fig:resl_mode} (a)) which is not directly linked to the frequency of the leading eigenmode from global stability analysis. This was also observed by Qadri and Schmid [\citen{Qadri:PRF17}] suggesting that the difference in the peaks is caused by the non-normality of the linear operator. As seen in Fig.~\ref{fig:resl_mode} (b) and (c), the optimal forcing and response modes show their presence around the cavity leading edge and trailing edge, respectively. We find that the spatial structures of both modes emerge in the shear-layer region, which signifies the importance of shear-layer physics in open cavity flow.

Thus far, the discussions on cavity flows have assumed spanwise periodicity.  However, the presence of sidewalls is known to influence the global characteristics of the cavity flows [\citen{Crook:JFM13, Ohmichi:AIAAJ14, Beresh:JFM16, Sun:AIAAJ19}]. The sidewall effects on open-cavity flows can be considered by performing full 3D simulations and triglobal analysis without any Fourier expansions in the spanwise direction [\citen{Liu:JFM16}].

\subsection{Flow control}
\label{cav_control}

Insights from modal analysis can serve as a useful tool for the design of active flow control techniques.  Here, we highlight some of the recent efforts on performing open- and closed-loop control of cavity flows based on modal analysis.  To suppress hydrodynamic and pressure oscillations in cavity flows with open-loop control, actuators can be placed along the leading edge of the cavity in a 2D or 3D arrangement.  Two-dimensional control setups (spanwise invariant setting) have been examined experimentally [\citen{Shaw:AIAA99,Ukeiley:AIAAJ04,Dudley2014}], but simultaneous suppression of all resonant tones remains a challenge. On the other hand, 3D actuation (spanwise varying) has been found to be effective in reducing amplitudes across all resonant tones [\citen{Ukeiley:JA08, Takahashi:JFST11, Zhang:AIAAJ19, Sun:AIAAJ19}].

\begin{figure}
   \centering
   \includegraphics[width=0.9\textwidth]{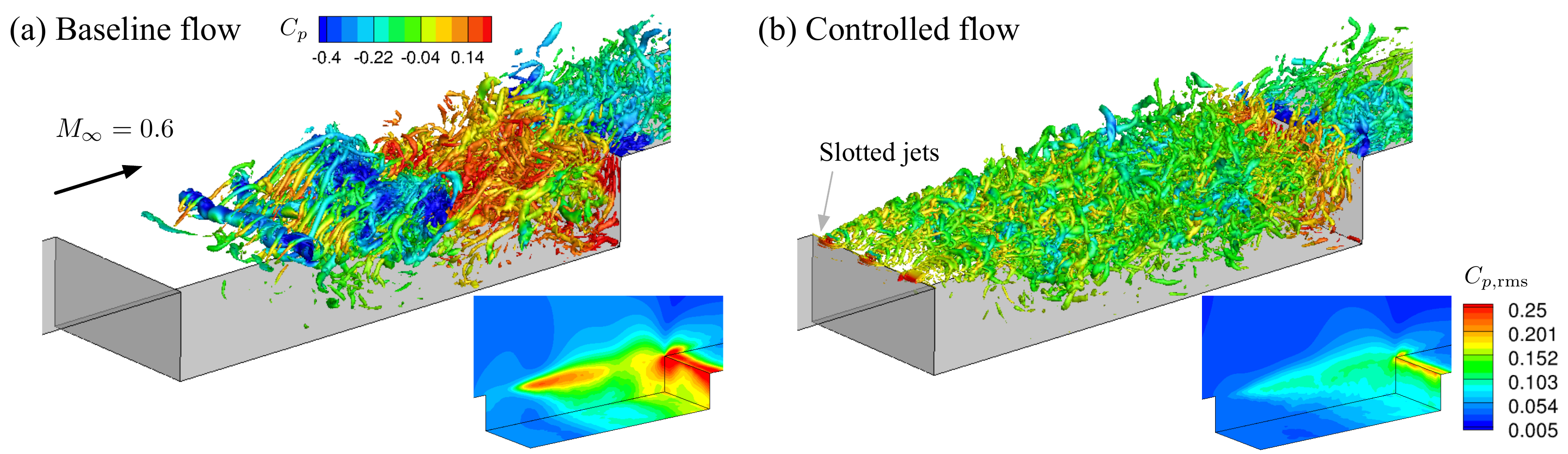}
   \caption{Active control of turbulent flow over a cavity of $L/D = 6$ at $Re_D = 10^4$ and $M_\infty = 0.6$ using 3D slotted-jets along the leading edge (figure adapted from [\citen{Sun:AIAAJ19}]).  $Q$ isosurface colored with the pressure coefficient $C_p$ is visualized over the cavity. Inserted plots show the pressure fluctuations. }
   \label{fig:Ctr_Sun}
\end{figure}

Modal analysis can help select the appropriate spanwise spacing between the actuators placed along the leading edge in a 3D setup.  That is, the preferred spanwise wavelength $\lambda$ (or wavenumber $\beta=2\pi/\lambda$) can be sought through the biglobal stability or resolvent analysis.  For example, Sun et al.~[\citen{Sun:AIAAJ19}] have used the insights from biglobal stability analysis to control turbulent flow over a spanwise periodic cavity of $L/D = 6$.  The goal of their work was to stimulate the emergence of 3D modes to remove kinetic energy from the dominant 2D shear-layer modes that are responsible for the large-amplitude fluctuations. The baseline and controlled flows from their study for $Re_D = 10^4$ and $M_\infty=0.6$ are shown in \fig \ref{fig:Ctr_Sun}.  The controlled flow uses a steady-jet actuation with the spanwise wavenumber that corresponds to the leading 3D stability mode, as reported above in \fig \ref{fig:BiG_Sun}.  While the stability analysis was performed for a much lower of $Re_D=502$, the insights from modal analysis appear to span across Reynolds number and effectively modify the turbulent cavity flow. As visualized in \fig \ref{fig:Ctr_Sun} with the $Q$-criterion [\citen{Hunt:CTR88}], spanwise coherent structures appearing in the baseline flow (around $x/D\approx2$) are inhibited in the controlled flow due to the 3D streaks introduced by the steady blowing. By preventing the spanwise shear-layer roll-ups, significant reduction in the levels of hydrodynamic and acoustic fluctuations are achieved as seen from root-mean-square of pressure displayed in subplots. The control strategy presented here has also shown its effectiveness in reducing pressure fluctuations for supersonic flows [\citen{George:AIAA15,Lusk:EF12}] as well as those for finite-span cavities [\citen{Sun:AIAAJ19}].

We can also use the spatial modes from modal analysis to reduce the state dimension of cavity flow and develop feedback (close-loop) control strategies. A reduced-order model of cavity flow can be obtained via projecting the state dynamics onto an appropriate set of modes. In the work of Barbagallo et al.~[\citen{Barbagallo:2009}], the global, POD, and BPOD modes were considered to design a close-loop control of an incompressible 2D open-cavity flow at $Re_D=7\,500$ to suppress flow unsteadiness.  An actuator and a sensor are placed on the walls and near the cavity leading and trailing edges, respectively. The actuator input was specified through wall-normal unsteady blowing/suction, while the output was taken to be the wall-normal shear stress integrated over the spanwise extent of the sensor. 

Barbagallo et al.~computed the global instability modes by solving the eigenvalue problems resulting from the linearized Navier--Stokes equations about the unstable steady state.  The POD and BPOD modes were determined by using the snapshots of the impulse response of the flow field.  Once the modes were obtained, the POD and BPOD modes were arranged in the order of the energy content and Hankel singular values, respectively, and the global modes were ordered according to their growth rates.  
The findings showed that the POD and BPOD modes performed well in modeling and suppressing the cavity flow oscillations.  The POD modes appear to exhibit robustness in the model, while the BPOD-based model is able to use far fewer modes compared to the POD based model to develop a feedback control law, because the BPOD modes effectively balance the controllability and observability of the system.  The reduced-order model constructed from global modes on the other hand over-emphasized the contributions from the highly damped modes to the input-output behavior.

\subsection{Landing gear well}

\begin{figure}
   \centering
   \includegraphics[width=0.85\textwidth]{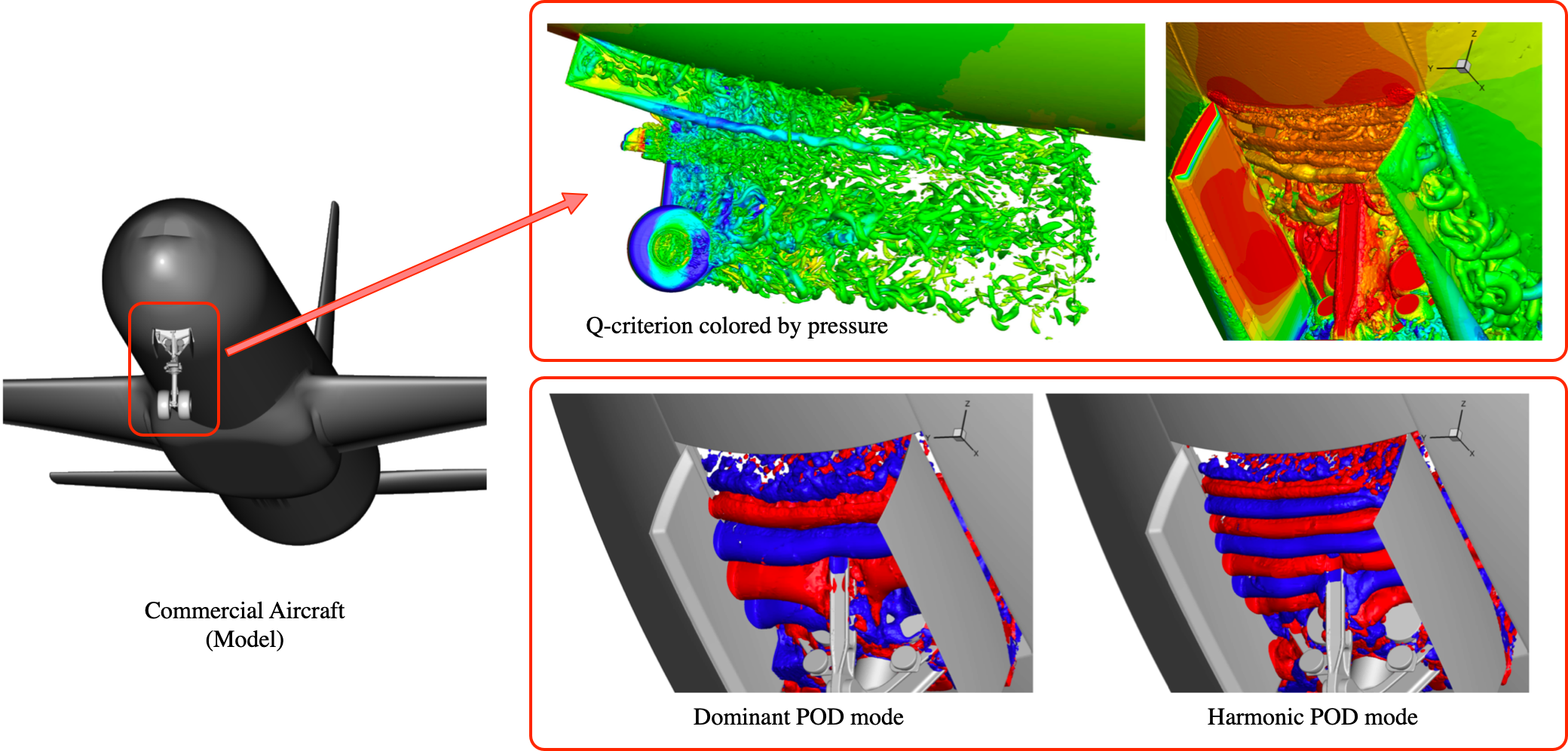}
   \caption{Turbulent flow over a model landing gear well of a commercial aircraft (top).   Transverse velocity POD mode captures the structures responsible for noise generation (bottom). Figure adapted from [\citen{Ricciardi:AIAA2019}] and reprinted with permission from the American Institute of Aeronautics and Astronautics.}
   \label{fig:landing_gear}
\end{figure}

We close this section on cavity flows by presenting an application of modal analysis to turbulent flow over a landing gear well of a commercial aircraft model by Ricciardi et al.~[\citen{Ricciardi:AIAA2019}].  The turbulent flow field obtained from a delayed detached-eddy simulation and the extracted dominant POD modes are shown in \fig \ref{fig:landing_gear}.  In this analysis, they identified the velocity POD modes [\citen{Sieber:JFM16}] that correlate with vortical structures responsible for generating the acoustic tones.  Although the turbulent flow over the landing-gear well is highly complex, the shear layer modes appear clearly, sharing great similarity with those from fundamental cavity flows.  This example highlights the importance of basic insights gained from basis modal analysis being beneficial for practical applications.

\section{Outlook}
\label{sec:outlook}

In this review, we have explored a number of modal analysis applications for fluid flows, ranging from simple, canonical flows to highly complex, real-world engineering configurations.  One key takeaway is that each flow has its own unique blend of challenges, including high-dimensionality, nonlinearity, multiscale phenomena, complex geometry, and non-normality.  In this outlook, we discuss challenges and limitations of existing methods, as well as emerging techniques in machine learning and data science to address these challenges for reduced-order modeling and control.  Below, we summarize several important avenues of ongoing research that are likely to have significant impact on modal analysis in fluids.

\subsection{Beyond linear superposition and separation of variables}

Most of the modal extraction techniques discussed above have been fundamentally linear, relying on superposition.  For example, the POD and DMD modal bases both approximate the flow field in terms of a \emph{linear} combination of these modes.  However, we know that many flows evolve on a low-dimensional manifold [\citen{Noack:JFM03}], rather than a low-dimensional linear subspace.  Thus, there is potential to obtain more accurate and efficient reduced order models with \emph{nonlinear} dimensionality reduction techniques.  
There are a number of effective techniques, for example based on locally linear embedding~[\citen{roweis2000nonlinear,ehlert2019locally}] and Grassmannian manifolds~[\citen{amsallem2008interpolation,Loiseau2019arxiv}]. 
In machine learning, the autoencoder is a neural network that embeds high-dimensional data into a low-dimensional latent space, followed by a decoder network that lifts back to the ambient high-dimensional space.  Linear autoencoders have been shown to identify reduced subspaces similar to those found with POD, and they may be generalized to develop nonlinear analogues of POD~[\citen{Milano2002jcp,MurataXX}].  Moreover, it has been shown that sufficiently large networks can represent arbitrarily complex input--output functions~[\citen{hornik1989multilayer}].  For this reason, {\it deep} network architectures [\citen{Krizhevsky2012nips, Lecun2015nature, Goodfellow-et-al-2016}] are becoming increasingly useful for modeling fluids, and dynamical systems more generally.  For example, deep learning has been recently used to identify nonlinear embeddings where the dynamics are linear [\citen{Wehmeyer2018jcp, Mardt2018natcomm, Takeishi2017nips, Yeung2017arxiv, Otto:SIAM2019, Lusch2018natcomm}], inspired by Koopman theory~[\citen{Mezic2005nd,Mezic:ARFM13,Brunton2016plosone}].  However, note that deep learning requires extremely large volumes of training data, and resulting models are generally only useful for interpolation.  

Snapshot-based dimensionality reduction methods that use the singular value decomposition, such as POD and DMD, are fundamentally based on a space-time separation of variables.  This can be seen clearly from the POD expansion in Eq.~\eqref{eq:PODexpansion}.  However, many problems in fluid mechanics involve traveling wave phenomena that are not well modeled by separation of variables.  Simple symmetry transformations of a coherent structure, such as translations and rotations, will result in a large number of POD modes~[\citen{Brunton2019book}].  Moving away from this space-time separation of variables is one of the most important open challenges in modal analysis, with the potential to improve how we describe and model convecting coherent structures.

\subsection{Sparse and randomized algorithms}

Dimensionality reduction and sparse algorithms are synergistic, in that underlying low-dimensional representations facilitate sparse measurements~[\citen{Manohar2017csm,Mohren2018pnas}] and fast randomized computations~[\citen{Halko2011siamreview}].  Compressed sensing has already been leveraged for compact representations of wall-bounded turbulence~[\citen{Bourguignon2014pof}] and for POD based flow reconstruction~[\citen{Bai2014aiaa}].  If only classification or detection is required, reconstruction can be circumvented and the measurements can become orders-of-magnitude sparser [\citen{Bright2013pof, Kramer2017siads, Brunton2016siap}].  Decreasing the amount of data to train and execute a model is important when a fast decision is required, as in control. 

Low-dimensional structure in data also facilitates acceleration of computations via randomized linear algebra [\citen{Mahoney2011,Halko2011siamreview}].  If a matrix has low-rank structure, there are efficient matrix decomposition algorithms based on random sampling that can be adopted.  
The basic idea is that if a large matrix has low-dimensional structure, then with high probability this structure will be preserved after randomly projecting the columns or rows onto a low-dimensional subspace, facilitating efficient computations.  These so-called \emph{randomized} numerical methods have the potential to transform computational linear algebra, providing  accurate matrix decompositions at a fraction of the cost of deterministic methods.  For example, randomized linear algebra may be used to efficiently compute the singular value decomposition.  Randomized POD [\citen{rokhlin2009randomized, halko2011algorithm}], randomized DMD [\citen{Bistrian2016ijnme, Erichson2017arxiv,hemati2015tls}], and randomized resolvent analysis [\citen{ribeiro2019randomized, Moarref13}] have also been developed based on the same principles.  

\subsection{Machine learning for reduced-order models and closures}

Beyond a detailed analysis and characterization of fluid flows, one of the overarching goals of modal analysis is the construction of {\it predictive} models that may be used for design, optimization, estimation, and control. One of the primary challenges for effective flow control is the computational complexity and latency associated with making a control decision, which may introduce unacceptable time-delays and destroy robust control performance [\citen{Brunton2015amr}]. Fully resolved simulations of multiscale flow phenomena are generally too slow for real-time feedback control.  Thus, significant effort has gone into developing ROMs, with the goal of accurately and efficiently reproducing only the most relevant flow mechanisms~[\citen{Benner2015siamreview,Rowley:ARFM17}].  A classical approach involves Galerkin projection of the governing Navier-Stokes equations onto an orthogonal basis, such as Fourier or POD modes, results in a system of ordinary differential equations for the mode coefficients. Alternative data-driven approaches may be used to develop reduced-order models via \emph{system identification}.  In both cases, machine learning techniques are emerging to improve the modal basis and models.  For instance, deep feedforward neural networks can be combined with POD-based modal analysis techniques to develop accurate reduced-order models even for high Reynolds number flows [\citen{Lui:JFM2019}].

Increasingly, machine learning is being used directly to build reduced-order models of physical systems from data.  These approaches may be broadly categorized into methods that identify self-contained models~[\citen{Glaz2010aiaa, Williams2015jcd, Williams2015jnls, Semeraro:PRF2017, Brunton2016pnas, Loiseau2017jfm, Loiseau2018jfm, wan2018data, vlachas2018data, raissi2018hidden, raissi2019physics}] and methods that develop closures for existing coarse-grained models, such as POD-Galerkin [\citen{mohebujjaman2018physically}], RANS [\citen{Ling2015pf, Parish2016jcp, Ling2016jfm, Xiao2016jcp, Singh2017aiaaj, Wang2017prf}], and LES [\citen{Maulik2019jfm}]; for an excellent review of data-driven closure models, see [\citen{kdur_ageofdata}].  It is also important to distinguish what types of input data are required to construct a reduced-order model, including resolution in space and time, volume and quality of data, and whether or not experimentally inaccessible information, such as from an adjoint simulation, is required.  

Because fluid flow modeling is central to many applications in health, security, and transportation, it is often essential that models be interpretable.  It is no surprise that these are among the leading challenges in machine learning and artificial intelligence research.  The data-driven modeling of fluid flows is a rapidly growing field. Thus, we provide a high-level summary of some representative examples.  

\subsubsection{Dynamical systems models} 

There are several approaches for the modeling of time-series data that have been applied to fluid flow systems.  Neural networks are often used for nonlinear system identification, as in NARMAX~[\citen{Semeraro:PRF2017,Glaz2010aiaa}].  Long short-term memory (LSTM) networks, which have been widely applied for speech recognition, are now being used to model chaotic dynamical systems~[\citen{wan2018data,vlachas2018data}].  Deep learning is also being broadly used to model systems in physics~[\citen{raissi2018hidden,raissi2019physics}].  Kernel methods have been employed to enrich the space of measurement functions used to approximate the Koopman operator via the \emph{extended DMD}~[\citen{Williams2015jcd,Williams2015jnls}].  However, neural networks and kernel methods typically result in black-box models that may be prone to overfitting, unless care is taken to cross-validate results.  In modeling dynamical systems, the principle of \emph{parsimony} states that a model should have the lowest complexity possible, while still faithfully representing the observed phenomena.  That is, the balance between \emph{model complexity} and \emph{model misfit} is of central importance.  This balance is helpful for preventing overfitting and promoting models that are interpretable---since there are only a few terms in the model that may be connected to physical interactions---and generalizable~[\citen{Brunton2016pnas}].  The sparse identification of nonlinear dynamics (SINDy) method identifies the fewest terms required to model time-series data with a differential equation, and has recently been applied to model various fluid flows [\citen{Loiseau2017jfm, Rudy2017sciadv, Loiseau2018jfm}].  By building models directly on physically intrinsic quantities, such as lift and drag measurements, these models bypass the well-known challenges of projection based methods of continuous mode deformation associated with changing geometry and flow conditions [\citen{Loiseau2018jfm}].  Similar approaches have also been considered from the standpoint of prediction-error and subspace identification techniques for parameter-varying models [\citen{hemati2016aiaa}], albeit without the advantage of model interpretability.  It is also possible to combine modal analysis with the theory of networked dynamical systems [\citen{Newman06, Newman10,Kaiser2014jfm,Nair:JFM2019,Nair:JFM15,Taira2016jfm,Nair:PRE18,meena2018network}].  Network-based approaches have been used in several novel flow applications~[\citen{Nair:JFM15}], including to obtain cluster reduced-order models of complex flows~[\citen{Kaiser2014jfm,Nair:JFM2019}], to model two-dimensional isotropic turbulence~[\citen{Taira2016jfm}], to model and control wake flows~[\citen{Nair:PRE18}], and for community detection in wake flows~[\citen{meena2018network}].  Finally, many of the methods above, including DMD with control~[\citen{Proctor:2016DMDc}], extended DMD~[\citen{williams2014edmd}], streaming DMD~[\citen{hemati2014streaming,hematistdmd2016,zhangArxiv2017}], and SINDy, have been used in conjunction with model predictive control to tame complex dynamical systems and fluid flows~[\citen{Korda2018automatica,Kaiser2018prsa,arbabi2018data,deem2018}].
 
\subsubsection{Closure models and stabilization} 

Many classical reduced-order modeling approaches involve truncating the modal basis to only include large, dominant, energetic coherent structures, neglecting the detailed modeling of fine-scale structures.  However, even if the truncated modes do not contain a significant portion of the energy, they can still play a significant role in the dynamics and stability of the ROM~[\citen{Wang_ROM_thesis}].  A variety of strategies have been developed that compute appropriate projections to ensure stability.  These efforts include energy-based inner products~[\citen{Kalashnikova_sand2014,rowley_pod_energyproj}], symmetry transformations~[\citen{Sirovich1987}], and the least-squares Petrov-Galerkin (LSPG) approach~[\citen{Carlberg2011ijnme}].  
From a practical standpoint, the LSPG method has to be employed with hyper-reduction, which will be discussed in the next section, since without hyper-reduction, the  computational cost of LSPG may be higher than that of the original full-order model. 
From a closure viewpoint, inaccuracies, which can lead to instabilities, are considered to result from the impact of the unresolved physics on the resolved modes.  Indeed, the error evolution is related to the closure terms, which is the target of subgrid-scale modeling in large eddy simulations.  Research has examined the construction of mixing length~[\citen{Aubry:JFM88}], Smagorinsky-type~[\citen{Ullmann_smag,wang_smag,smag_ROM,Wang_ROM_thesis}], and variational multiscale (VMS) closures~[\citen{san_iliescu_geostrophic,Wang_ROM_thesis,Bergmann_pod_vms}].  Another approach that displays similarities to the variational multiscale method is the Mori--Zwanzig (MZ) formalism~[\citen{StatisticalMechanics,stinis_finitememory,parishMZ1,parish_dtau,GouasmiMZ1,parishVMS}].  In VMS and MZ-based approaches, the state variables are decomposed into a resolved  set and an unresolved  set. The impact of the unresolved scales on the resolved scales is then modeled. Parish et al.~[\citen{parish2018residual}] represent and model unresolved physics in the form of a  memory integral that depends on the temporal history of the coarse-scale variables. This approach presents a unified view of VMS closures and Petrov--Galerkin stabilization.

\subsubsection{Hyper-reduction} 

Even though the reduced-order model equations are in terms of a reduced state of dimension $r$, the projected dynamics require evaluation of the high-dimensional nonlinear dynamics, which are of dimension $n\gg r$.  This limits the utility of ROMs of nonlinear systems, as the online cost can scale as $O(n)$. Thus, beyond reducing the order of a model, acceleration (or alternately \emph{hyper-reduction} or \emph{sparse sampling}) techniques will be required to improve the efficiency of a ROM.  The gappy POD method provides the ability to sparsely sample a system in $O(r)$ locations and still evaluate the POD and terms in the Galerkin projection [\citen{Everson1995josaa,willcox2006unsteady}].  In addition, there are reduced-basis methods for PDEs [\citen{Barrault2004crm}] and the associated discrete empirical interpolation method (DEIM) [\citen{Chaturantabut2010siamjsc, Chaturantabut2012siamjna, peherstorfer2014localized}], which approximates nonlinear terms by evaluating the nonlinearity at a few points.  This approach is also prevalent in high-performance computing [\citen{Carlberg2011ijnme, Avellaneda1990cmp, Amsallem2012ijnme, Carlberg2013jcp}], and may also impact flow control.


\section{Conclusions}

As a sequel to the previous introductory overview paper on modal analysis [\citen{Taira_etal:AIAAJ17}], this document surveyed applications of modal analysis techniques with the hope to serve as a {\it go-to} guide for readers seeking insights on how modal analysis techniques can help analyze different types of flows.  With such point in mind, the present paper focused on presenting applications of modal analysis techniques to study, model, and control canonical aerodynamic flows.  To illustrate how modal analysis techniques can provide physical insights in a complementary manner, we selected four fundamental examples of cylinder wakes, wall-bounded flows, airfoil wakes, and cavity flows.  A good portion of the examples considered in this paper considered the applications of modal analysis for developing effective active flow control strategies.   While this paper attempted to cover a range of topics, it is by no means comprehensive in nature.  For readers with elevated levels of interests, we invite them to delve into the references for details.

Towards the end of this paper, we also offered some brief discussions on the outlook for modal analysis techniques, in light of rapid developments in data science.  As we see emergence of many refreshing data-inspired concepts, we fluid mechanicians are in an exciting era to incorporate these ideas and extend modal analysis techniques.  In fact, there are ongoing developments in handling large data sets and constructing sparse interpretable models.  We believe such efforts can enable the analysis of high-dimensional fluid flows with complex dynamics.  It is also noteworthy that the implementation of these approaches are facilitated with the enhancement in the available computational resources.  We hope that this overview paper, along with the first overview paper, serves as a valuable educational tool for engineers and scientists interested in performing modal analysis of aerodynamic flows.


\section*{Acknowledgments}

This paper was one of the major outcomes from the AIAA Discussion Group (Fluid Dynamics Technical Committee) entitled ``Modal Decomposition of Aerodynamic Flows'' organized by Douglas Smith, Kunihiko Taira, Maziar Hemati, and Karthik Duraisamy.  The authors thank the fruitful discussions with the members of the discussion group and greatly acknowledge the generous support from the following agencies: 
Air Force Office of Scientific Research~(AFOSR), 
Army Research Office~(ARO), 
National Science Foundation~(NSF),  
Office of Naval Research~(ONR), and Defense Advanced Research Projects Agency (DARPA). 
At last, we sincerely thank the Editor-in-Chief, Prof.~Alexander Smits, and Associate Editor, Prof.~Peyman Givi, of the AIAA Journal for supporting the overall effort to make this special section on Modal Analysis possible.


\bibliographystyle{aiaa}
\bibliography{refs_all,refs_channel}

\end{document}